\documentclass[aps,prd,twocolumn,amsmath,groupedaddress,nofootinbib]{revtex4}

\usepackage{graphicx}
\usepackage{amssymb}
\usepackage{bm}
\usepackage{dcolumn}

\usepackage{color}   

\newcommand{\be}{\begin{eqnarray}}      \newcommand{\ee}{\end{eqnarray}}   
\newcommand{\ba}{\begin{array}}         \newcommand{\ea}{\end{array}} 
\newcommand{\bs}{\begin{subequations}}  \newcommand{\es}{\end{subequations}} 
\newcommand{\rf}[1]{~(\ref{#1})}  
\newcommand{\rfs}[2]{~(\ref{#1})-(\ref{#2})}

\newcommand{\pg}[1]{p.~\pageref{#1}} 
\newcommand{\ct}[1]{~\cite{#1}}  
\newcommand{\lb}[1]{\label{#1}}  

\newcommand{\nn}{\nonumber}
\newcommand{\lf}{\left}     \newcommand{\rt}{\right}
\newcommand{\fr}{\frac}     

\newcommand{\ov}{\over}    
\newcommand{\la}{\langle}    \newcommand{\ra}{\rangle}     
\newcommand{\nbrk}[1]{\mbox{$#1$}}

\newcommand{\Vsp}{\vphantom{\displaystyle{\hat I \over \hat I}}}
\newcommand{\vsp}{\vphantom{\displaystyle{I \over I}}}
\newcommand{\rbx}[1]{\raisebox{1.5ex}[0pt]{#1}}

\def\d{\delta}    \def\pd{\partial}  \def\D{\Delta}
\def\dd{\d_{\rm D}}
\def\dc{\delta^{(c)}}
\def\t{\tau}       \def\s{\sigma}    
\def\al{\alpha}    \def\b{\beta}     \def\g{\gamma}
\def\l{\lambda} 
\def\Nb{\nabla}   \def\Nbi{\nabla_{\!i}}    \def\Nbj{\nabla_{\!j}}

\def\f{\varphi}         

\def\Om{\Omega}        
\def\eps{\epsilon} 
\def\H{{\cal H}}
\def\R{{\cal R}}
\def\Teff{\Theta_{\rm eff}}    
\def\eff{_{\rm eff}}

\def\di{\iota}
\def\dip{q}
\def\x{{\bm x}}      \def\r{{\bm r}}         
\def\k{{\bm k}}      \def\p{{\bm p}}      \def\q{{\bm q}} 
\def\^{\hat}            \def\~{\tilde}  
\def\n{\hat{\bf n}} 
\def\en{\hat{\bm{\epsilon}}}     
\def\odot{^{\!\displaystyle\cdot}}

\def\snu{_{\nu}}           \def\sg{_{\gamma}}  
\def\spr{_{\rm pr}}

\def\rtg{\sqrt{-g}\,}
\def\dOmn{d^2n_i}  
\def\pip{\pi_{\!p}}     \def\Pip{\Pi_{\!p}} 
\def\eg{{\it e.g.\/}}   \def\cf{{\it c.f.\/}\ }   \def\ie{{\it i.e.\/}}

\def\i{_{\rm in}}

\newcommand{\const}{{\rm const}}

\newcommand{\diag}{\mathop{\rm diag}}

\def\cmbf{{\sc cmbfast}}

\begin{document}


\title{\Large  Gravity of Cosmological Perturbations 
               in the CMB}

\author{Sergei Bashinsky}

\affiliation{Theoretical Division, T-8, 
             Los Alamos National Laboratory, Los Alamos, NM 87545, USA}
\affiliation{International Centre for Theoretical Physics, 
          Strada Costiera 11, Trieste, Italy}

\date{\today}
\date{October 4, 2006}

\begin{abstract}

   First, we establish which measures of large-scale perturbations
are least afflicted by gauge artifacts and directly map the apparent evolution 
of inhomogeneities to local interactions of cosmological species. 
Considering nonlinear and linear perturbations of phase-space distribution, 
radiation intensity and arbitrary species' density, we require that: 
(i) the dynamics of perturbations defined by these measures
is determined by observables within the local Hubble volume;
(ii) the measures are practically applicable on microscopic scales and 
in an unperturbed geometry retain their microscopic meaning on all scales. 
We prove that all measures of linear overdensity 
that satisfy (i) and~(ii) coincide in the superhorizon limit.
Their dynamical equations are simpler than the traditional ones, 
have a nonsingular superhorizon limit and explicit Cauchy form.
Then we show that, contrary to the popular view,
the perturbations of the cosmic microwave background (CMB) in the radiation era 
are not resonantly boosted self-gravitationally during horizon entry. 
(Consequently, the CMB signatures of
uncoupled species which may be abundant in the radiation era, 
e.g. neutrinos or early quintessence, are mild;
albeit non-degenerate and robust to cosmic variance.) 
On the other hand, dark matter perturbations in the matter era
gravitationally suppress large-angle CMB anisotropy by an order of magnitude stronger 
than presently believed. If cold dark matter were the only dominant component 
then, for adiabatic perturbations, the CMB temperature power spectrum 
$C_\ell$ would be suppressed 25-fold.

\end{abstract}

\maketitle


\tableofcontents

\section{Introduction}
\lb{sec_intro}

Cosmological observations can be valuable probes
of matter species whose interactions
are too feeble to be studied by more traditional 
particle physics techniques.
Whenever such ``dark'' species, including dark energy, 
dark matter, or neutrinos, constitute a non-negligible
energy fraction of the universe, 
these species leave gravitational imprints on the distribution 
of more readily observable matter, notably,
the cosmic microwave background (CMB) and 
baryonic astrophysical objects.
In the framework of a perturbed cosmological expansion,
the dark species influence the visible matter
by affecting both the expansion rate, 
controlled by their average energy,
and the metric inhomogeneities,
sourced by species' perturbations.

The gravitational signatures of the perturbations
are particularly informative
as they are sensitive to
the internal (kinetic) properties of the dark species.
These signatures are generally dissimilar 
for species with identical background energy and pressure 
but different dynamics of perturbations,
\eg, for self-interacting particles, uncoupled particles, 
or classical fields\ct{HuEisenTegWhite98,Erickson01,BeanDore03,BS,Abramo04}. 
Perturbations of dark species can affect
the observed CMB and matter power spectra 
by an order of magnitude. 
For example, the gravitational impact 
of dark matter perturbations 
suppresses the CMB power at low multipoles $\ell\lesssim 200$ 
by up to a factor of 25 (Sec.~\ref{sec_mat}).

Despite the value and prominence of the gravitational impact
of perturbations, most of the existing descriptions of
cosmological evolution are misleading
in matching 
dynamics of perturbations in dark sectors to
features in observable distributions.
One long-investigated source of ambiguities
and erroneous conclusions
is the dependence of the apparent evolution of
perturbations and the induced by them gravitational fields
on the choice of spacetime coordinates (metric gauge), \eg\ct{Bardeen80,KS84}.
Of course, the CMB and large scale structure
observables should be identical in any gauge.
However, in various gauges 
the features of the observable distributions 
may appear to be generated 
in different cosmological epochs
and by different mechanisms.

For example, given the standard adiabatic initial conditions,
the perturbation of the CMB temperature~$\d T/T$
grows monotonically on superhorizon scales 
in the historically important and still popular 
synchronous gauge, \eg\ct{LandauLifshitzII};
$\d T/T$ generally remains constant beyond 
the horizon in a single cosmological epoch but 
changes during the transition to another epoch  
in the calculationally convenient and intuitive Newtonian gauge,
\eg\ct{Mukh_Rept,MaBert95};
or, $\d T/T$ is strictly frozen beyond the horizon 
but evolves differently since horizon entry 
in the spatially flat gauge.
These descriptions naively suggest
different separation of the presently measured
CMB temperature anisotropy 
into the inhomogeneities of primordial (inflationary) origin
and those generated by the
gravitational impact of perturbations in
various species (CMB, neutrinos, 
dark matter, dark energy, etc.)


The use of gauge-invariant perturbation variables\footnote{
  A ``gauge-invariant perturbation variable'' 
  generally cannot be measured by a local observer.
  (For example, the local values of the gauge-invariant
   Bardeen potentials~$\Phi$ and $\Psi$\ct{Bardeen80}
   are determined by spacetime curvature 
   outside the observer's past light cone.)
  Thus the gauge-invariant perturbations 
  should be distinguished from gauge-invariant 
  local observables or from local  
  physical tensor quantities, such as the energy-momentum tensor~$T^{\mu\nu}$ 
  or the curvature tensor~$R^{\mu}_{\nu\al\beta}$.
}\ct{GerlachSengupta78,Bardeen80,KS84,BruniDunsbyEllis92}
does not remove this ambiguity\ct{Unruh98}. 
Indeed, the perturbations of species' density and velocity  
or the metric 
in {\it any\/} fixed gauge can be written as gauge-invariant expressions,
which may be called ``gauge-invariant'' (though, clearly, non-unique) 
definitions of these perturbations\ct{Unruh98}.  
Hence, the variety of gauge-invariant 
perturbation variables is at least as large  
as the variety of gauge-fixing methods.

Nevertheless, without contradicting general covariance,
much of this descriptional ambiguity can be avoided.
We show that in cosmological applications 
the perturbed evolution
can be described  by variables whose change 
is necessarily induced by a local physical cause.
Moreover, when the equations
of perturbation dynamics are presented in terms of such variables,
the structure of the equations simplifies considerably.
The simpler equations allow tractable analytical analysis of
perturbation evolution for realistic cosmological models.

We impose two requirements 
on a measure of dynamical cosmological perturbations:
\begin{enumerate}
\item[I.]
The dynamics of the measure is determined completely by
locally identifiable physical phenomena within 
the local Hubble volume;
\item[II.]
The measure is universally applicable to all scales:
from superhorizon (governed by general relativity)
to subhorizon (governed by microscopic kinetics).
\end{enumerate}The goals of this paper are, 
first (Secs.~\ref{sec_form} and~\ref{sec_newt_dyn}), 
to provide a full dynamical description of cosmological inhomogeneities 
in terms of measures
which satisfy these two requirements.
Second (Secs.~\ref{sec_evol} and~\ref{sec_CMBfeatures}), 
to establish the origin of features in the CMB angular spectrum,
as revealed by this more direct description, 
and to explore the utility of these features for
probing the dark sectors.

Most of the best-known formalisms for the dynamics
of CMB and matter inhomogeneities are formulated in terms of
perturbations of physical observables,
such as radiation temperature $T(x^\mu,\n)$ or 
proper energy density~$\rho(x^\mu)$\ct{BondEfstathiou84,MaBert95,
CMBFAST96,ZalSel_allsky97,Dod_book}.
When these perturbations are evaluated in the Newtonian gauge,
nonsingular on small scales,
then the small-scale dynamics in these formalisms does reduce to
special-relativistic kinetics and Newtonian gravity.
Then, however, the perturbation evolution 
beyond the Hubble scale is nonlocal\footnote{
  For illustration, 
a perturbation of energy density~$\rho$ in a dust (zero pressure) universe
evolves in the Newtonian gauge\rf{Newt_gauge_def} as
$$
\d\dot\rho + 3\fr{\dot a}{a}\,\d\rho
    + \rho \Nbi v^i = 3\rho\,\dot\Phi,
$$
with
$$
\Phi=4\pi Ga^2\fr1{\Nb^2}\lf[\d\rho-3\fr{\dot a}{a}\,\rho\fr1{\Nb^2}\Nbi v^i\rt].
$$
Although the first equation above is obtained by 
linearization of a causal relation~$T^{0\mu}_{~;\mu}=0$,
the operators $1/{\Nb^2}$ in the second equation 
signal that after elimination of $\Phi$
the evolution of $\d\rho$ becomes nonlocal.
This does not violate causality because $\d\rho$
cannot be measured locally.
} 
and subject to gauge artifacts.
The artifacts are caused by the nonlocality 
of defining the motion of coordinate observers, 
hence, inferring the components of tensors
and the splitting of dynamical variables into their 
background values and perturbations from 
the gauge conditions.
While some gauge conditions, \eg\ synchronous, 
can be imposed locally, the resulting descriptions
tend to be singular, contain gauge modes,
and fail to reduce to Newtonian gravity
on small scales.

Studies of the connection of the observed
cosmological inhomogeneities to 
quantum fluctuations of an inflaton field during inflation
have led to an extensive list of variables which under certain 
conditions freeze (become constant in time) beyond the horizon.
The best known examples are the Bardeen 
curvature~$\zeta$\ct{Bardeen80,zeta_orig} (conventionally interpreted 
either as a perturbation of intrinsic curvature 
on uniform-density hypersurfaces
or as density perturbation on spatially flat hypersurfaces) 
and the curvature perturbation~$\R$ 
on comoving hypersurfaces, \eg\ct{Lyth84}. 
Both $\zeta$ and~$\R$ are frozen for adiabatic superhorizon perturbations.
For the general, nonadiabatic, superhorizon perturbations, 
the curvature perturbations~$\zeta_a$\ct{zeta_a} on the 
hypersurfaces of uniform energy density of 
an individual minimally coupled 
perfect fluid~$a$ were shown to be also frozen (``conserved'').\footnote{
  The perturbations~$\zeta_a$\ct{zeta_a}
  are conserved for uncoupled perfect fluids\ct{zeta_a} 
  or species whose perturbations are internally adiabatic\ct{BS} 
  only if {\it gravitational\/} decays\ct{LindeMukh_curvaton05} 
  of species into another type of species are negligible. 
  This should be a reasonable assumption for realistic applications
  to the evolution of the CMB and large scale structure.
} 
Finally, perturbed cosmological evolution has been described
in terms of conserved spatial gradients 
of various quantities\ct{EllisBruni89,RigopoulosShellard03,Langlois05}.
The evolution of any of these variables is 
more robust to gauge artifacts on superhorizon scales than,
for example, that of $\d T/T$, $\d\rho/\rho$, or 
(the Bardeen) potentials $\Phi$ and $\Psi$ in the Newtonian gauge.
However, neither the uniform-density, nor comoving, 
nor spatially flat hypersurfaces
reduce to Newtonian hypersurfaces on subhorizon scales. 
Nor does the evolution of gradients tend to the familiar
picture of particles and fields propagating in Minkowski spacetime.

In Sec.~\ref{sec_form}
we consider natural measures for 
nonlinear perturbations of phase-space distribution 
and radiation intensity 
and for linear perturbations of species' density
that conform to both requirements I and~II.
Moreover, we prove in Sec.~\ref{sec_form_mapping}
that, although there is no ``physically preferred''
description of perturbed evolution, 
the overall change of density perturbations
of any species during horizon entry is the same in any
description which satisfies certain conditions
formalizing requirements I and~II. 
This change is different in most of 
the traditional formalisms, violating these requirements.

In Sec.~\ref{sec_newt_dyn} we describe a closed 
formulation of linear dynamics of scalar perturbations
in terms of the above measures evaluated in the Newtonian gauge.
We find that this formulation has several technical 
advantages, ultimately related to its tighter connection
between the perturbation measures and 
the causality of the perturbed cosmological dynamics. The proposed approach 
has broad scope of applicability, providing an economical and physically 
adequate description of phenomena which involve inhomogeneous evolution 
of multiple species and non-negligible general-relativistic effects. 
Examples include the physics of inflation, reheating,
the CMB, and cosmic structure. In this paper we focus on the last two topics.

After a concise review 
of the evolution of perturbations
on superhorizon scales and during horizon entry in Sec.~\ref{sec_evol},
we apply in Sec.~\ref{sec_CMBfeatures}
the developed formalism to study
the gravitational signatures of various
species in CMB temperature anisotropy and large scale structure.

Contrary to a popular view,
\eg\ct{HuSugSmall95,HuNature,HuFukZalTeg00,HuDodelson02,Dod_book,Mukhanov_book}, 
based on the study of traditional proper perturbations 
in the Newtonian gauge, 
we find that
CMB perturbations in the radiation era are not
resonantly boosted by their self-gravity (Sec.~\ref{sec_rad_driv}).
As a consequence, the gravitational signatures 
of dark species in the radiation era,
such as neutrinos or a dynamical scalar field 
(quintessence\ct{Wetterich88,RatraPeebles88}),
are moderate; as these species, even when they are abundant, 
do not untune any physical resonant amplification.
Fortunately, due to low 
cosmic variance on the corresponding scales and the existence 
of characteristic nondegenerate signatures (Sec.~\ref{sec_rad_impact}), 
we can still expect meaningful robust constraints on the nature of 
the dark radiation sector.

On the other hand, the gravitational impact 
of dark sectors' perturbations on the CMB in the matter era  
is found to be an order of magnitude 
stronger than the traditional interpretations suggest (Sec.~\ref{sec_mat}). 

%
%
\begin{table*}[t]
\begin{tabular}{|clc|}
\hline
Symbol & \qquad\qquad\qquad\qquad  Meaning  &    Definition \\
\hline
$\t$  &  Coordinate time, $x^0$ [Conformal time in the FRW background]  &  
                        Sec.~\ref{sec_phase_sp} [eq.\rf{Newt_gauge_def}]    \\
$\x$  &  Spatial coordinates, $x^i$ [Comoving coordinates in the FRW background] 
                    & \phantom{Sec.~\ref{sec_phase_sp}} [eq.\rf{Newt_gauge_def}] \\
$P_i$ &  Canonical momenta  &  Sec.~\ref{sec_phase_sp} \\
$P$   &  $(\sum_{i=1}^3 P_i^2)^{1/2}$ in any metric  
                                                &  eq.\rf{Pn_def_a} \\
$n_i$ &  Direction of propagation, $P_i/P$    &  eq.\rf{Pn_def_a} \\
$f(x^\mu,P_i)$   &  Phase-space distribution    & Sec.~\ref{sec_phase_sp} \\  
$df(x^i,P_i)$    &  Canonical perturbation of $f$, $f(x^i,P_i)-\bar f(P)$ 
                         &  Sec.~\ref{sec_phase_sp}, eq.\rf{df_def} \\  
$I(x^\mu,n_i)$   &  Conformal intensity of radiation     &   eq.\rf{I_def_a} \\
$\di(x^\mu,n_i)$ &  Perturbation of radiation intensity, ${I}/{\bar I}-1$ 
                                                         &   eq.\rf{di_def} \\
$f_{\al\b}$, $I_{\al\b}$, $\di_{\al\b}$ &  Describe polarized photons  &  
                                                        Sec.~\ref{sec_phot_P}\\
$\rho(x^\mu)$, $p(x^\mu)$ &  Energy density and pressure of species  &  \pg{rho_def}\\
             &  Perturbation of species' coordinate number density 
                                            & eq.\rf{d_a_def} or\rf{d_a_Newt} \\
\rbx{$d(x^\mu)$}  &  (for a fluid of particles, $\d n_{\rm coo}/n_{\rm coo}$ with
                 $n_{{\rm coo}}=dN/d^{3}\x$)     & (\pg{d_a_interp}) \\
$v_i(x^\mu)$  &  Normal bulk velocity of species    &  eq.\rf{v_i_def} \\
$v'^i(x^\mu)$ &  Coordinate bulk velocity
                                              &  eq.\rf{v_a_coo_def} \\
$u(x^\mu)$ & Velocity potential of scalar perturbations, $v_i = -\Nbi u$
                                              &  eq.\rf{vp_def} \\
$\s(x^\mu)$ & Scalar potential of anisotropic stress
                                              & eq.\rf{sigma_def}\\
$d_l(x^\mu)$ & Scalar multipole potentials of $\di$ 
               (in particular, $d_0=d$, $d_1=u$, $d_2=\s$)  
                                              & eq.\rf{dl_def}\\
$p_l(x^\mu)$ & Scalar multipole potentials of photon polarization 
                                              &  eq.\rf{pl_def}\\
$\phi(x^\mu)$ & A classical scalar field (quintessence)    
                                              &  Sec.~\ref{sec_quint} \\
\hline
$g_{\mu\nu}$ & Metric tensor, with the signature $(-,+,+,+)$ & \\
$a$   &  Scale factor in the FRW background        &   eq.\rf{metric} \\
$h_{\mu\nu}$ &  General perturbation of the metric, $\d g_{\mu\nu}/a^2$  
                                              &   eq.\rf{metric}  \\
$D$ and $\epsilon$ &  General scalar perturbations of the spatial metric $g_{ij}$ &  
                                                      eq.\rf{g_ij_param} \\
$\Phi$ and $\Psi$  &  Scalar perturbations of the metric in the Newtonian gauge &    
                                                  eq.\rf{Newt_gauge_def}  \\
$\zeta_a$ &  Reduced curvature, 
            $D+\fr13\Nb^2\epsilon=-\fr14\,a^2(\Nb^{-2}){}^{(3)}\!R$,
            of 3-slices $\rho_a=\const$ 
                                                &  Sec.~\ref{sec_dens} \\
\hline
overdot, $\dot{}$ \vphantom{$\fr{\hat1}1$}   &  $\pd/\pd\t$ &  \\
$\H$  &  Coordinate expansion rate [conformal, in the FRW background] 
                                 & Sec.~\ref{sec_phase_sp} ~[$\dot a/a$]\\
$\g$  &  
         $4\pi Ga^2(\rho+p)$ in the FRW background  & eq.\rf{gamma_def}  \\
$\di\eff$ and $\Teff$  &  Effective intensity and temperature perturbations  & 
                                     eqs.\rf{dieff_def} and\rf{Teff_def}\\
$c_s$     &  Speed of sound in the photon-baryon plasma 
                                                   & eq.\rf{c_s_def} \\
$R_b$     &  Baryon to photon enthalpy ratio, 
                      $3\rho_b/(4\rho_{\g})$  & Sec.~\ref{sec_CMB_prelim} \\
$S$ &  Acoustic horizon, $\int_0^\t c_s d\t$ 
           (when $R_b\ll1$, $\t/\sqrt{3}$)  & Sec.~\ref{sec_rad_impact} \\
$\f$ &  Phase of acoustic oscillations, $kS(\t)$  &  eq.\rf{d_rad} \\
$\t_c$    &  Mean $\t$ of a photon collisionless flight   
                                                   & eq.\rf{t_c_def}   \\
$~~\lf.{\d T / T}\rt|_{\rm in}$  &
     Primordial superhorizon perturbation of CMB temperature, 
     $\lf.\fr13\,d_{\g}\rt|_{k\ll\H}$ & eq.\rf{dT_in} \\
$\D T/T$  &  Presently observed perturbation of CMB temperature  
                                          & eq.\rf{los_zeta} \\     
\hline
\end{tabular}
\caption{Summary of the main notations.}
\lb{tab_notations}
\end{table*}
%
%

We summarize our results in Sec.~\ref{sec_concl}.
Appendix~\ref{sec_dyn} presents
linear dynamical equations for scalar perturbations 
of typical cosmological species in terms of the suggested measures.
Appendix~\ref{sec_Cl} summarizes the formulas for 
the scalar transfer functions and angular power spectra
of CMB temperature and polarization.
The main notations to be used in this paper are listed 
in Table~\ref{tab_notations}.

\section{Measures of Perturbations}
\lb{sec_form}

\subsection{Phase-space distribution}
\lb{sec_phase_sp}

Our first example of a quantity whose dynamics
can be fully determined by physics in the local Hubble volume
is the one-particle phase-space distribution
of classical, possibly relativistic, point particles
[photons, neutrinos, or cold dark matter (CDM)] 
$f(\t,x^i,P_i)=dN/d^3x^id^3P_i$.\lb{f_def}
(By default, Latin indices range from 1 to~3
and Greek from 0 to~3.)
We consider $f$ as a function of the coordinate time~$\t\equiv x^0$,\lb{tau_def}  
spatial coordinates~$x^i$, and, crucially for the following results,
{\it canonically conjugate\/} momenta~$P_i$.\lb{P_i_def}
The distribution~$f$ evolves according to 
the Boltzmann equation
\be
\dot f + \fr{dx}{d\t}^i\fr{\pd f}{\pd x^i}
       + \fr{dP_i}{d\t}\,\fr{\pd f}{\pd P_i} = C
\lb{df_dot_genr}
\ee
(for a systematic formulation of kinetic theory 
in general relativity see, \eg,\ct{Stewart_kinetics}.)
If the particles interact only gravitationally,
their canonical momenta~$P_i$ coincide with 
the spatial covariant components of the particle
4-momenta~$P^\mu$: $P_i=g_{i\mu}P^\mu$,
where $g_{\mu\nu}$ is the metric tensor.
Then
$dx^i\!/{d\t} = P^i\!/P^0$,
$dP_i\!/{d\t}$ is given by the geodesic equation,
and $C$, accounting for two-particle collisions, 
vanishes.
Hence,
\be
\dot f + \fr{P^i}{P^0}\,\fr{\pd f}{\pd x^i}
       + \fr{g_{\mu\nu,i}P^{\mu}P^{\nu}}{2P^0}\,\fr{\pd f}{\pd P_i} = 0.
\lb{df_dot_free}
\ee
We stress that this equation and the following 
observation apply to the fully {\it nonlinear\/} 
general-relativistic dynamics.

Let us consider the minimally coupled particles 
in the perturbed spacetime, 
possibly populated by additional species.
Let us also suppose that in certain coordinates the spatial scale,~$\l$, 
of inhomogeneities in the particle distribution
and in the metric is much larger than 
the temporal scale of local cosmological expansion,
~$\H^{-1}$, 
where for nonlinear theory $\H \equiv \fr16\fr{d}{d\t}\ln(\det g_{ij})$.\lb{aH_def}
Then from eq.\rf{df_dot_free}, where both the second and third terms 
contain spatial gradients ($\pd f/\pd x^i$ and $g_{\mu\nu,i}$),
we see that $\dot f/f = O(\l^{-1})\ll \H$.
In the limit of superhorizon
inhomogeneities ($\l\H\to \infty$)
the phase-space distribution~$f(\t,x^i,P_i)$ 
of minimally coupled particles becomes time-independent (frozen).
Then its background value,~$\bar f$,
and perturbation,~$d f\equiv f(x^\mu,P_i)-\bar f(P)$,\lb{df_int}
are also frozen.

Conversely, an apparent temporal {\it change\/} of~$d f$ 
(at fixed~$x^i$ and $P_i$ in any regular gauge)
can always be attributed either to the gravity of physical 
inhomogeneities within the local Hubble volume
or to local non-gravitational interaction.

The majority of contemporary formalisms
for cosmological evolution in phase-space,
\eg\ct{BondSzalay83,MaBert95,CMBFAST96,LiddleLyth_book,Dod_book},
work not with the canonical momentum of particles~$P_i$ 
but rather with {\it proper momentum\/}~$\p$,
measured by  coordinate or by normal observers.
These formalisms typically consider the phase-space distribution 
as a function of $\q\equiv a\p$,
the proper momentum rescaled by the background scale factor.
The evolution of the corresponding perturbation 
$\d f(x^\mu,\q)\equiv f(x^\mu,\q)-\bar f(q)$,
as well as of perturbations of integrated proper densities and intensities,
depends on the nonlocal procedure for
defining the observer's frame from the gauge condition.
In typical fixed gauges $\d f(x^\mu,\q)$, unlike $d f(x^\mu,P_i)$, 
generally changes even in the absence of 
local physical inhomogeneity or local non-gravitational coupling.\footnote{
  For example, for linear perturbations
  in the Newtonian gauge\rf{Newt_gauge_def},
  $q_i=P_i/(1-\Psi)$\ct{MaBert95,LiddleLyth_book,Dod_book}.  
  Hence, 
  $$
  \d f(x^\mu,\q) =  d f(x^\mu,P_i) - \Psi P_i\fr{\pd \bar f}{\pd P_i}.
  $$
  While the first term on the right-hand side evolves causally
  on superhorizon scales, the Newtonian potential~$\Psi$ does not,
  and so, neither does~$\d f(x^\mu,\q)$.  
}

Thus the stated in the introduction criteria I and~II 
for a measure of cosmological inhomogeneities 
are fulfilled for a perturbation of one-particle phase-space 
distribution $f(x^\mu,P_i)$, 
provided $P_i$ are the particle canonical momenta.
A description in terms of phase-space distributions
is, however, unnecessarily detailed
for most cosmological applications.
Instead, it is more convenient to use
integrated characteristics of the species,
such as intensities (for radiation of photons 
or other ultrarelativistic species) 
or energy densities and momentum-averaged velocities 
(for arbitrary species, 
including non-relativistic particles, 
fluids, or classical fields).
In the next two subsections we discuss the general-relativistic measures
of perturbations in intensity and density  
that also satisfy criteria I and~II.

\subsection{Radiation intensity}
\lb{sec_form_intens}

We start by considering inhomogeneities in the intensity of 
any type of cosmological radiation, such as
CMB photons or relic neutrinos 
at the redshifts at which 
the kinetic energy of the particles dominates their mass.
We describe the direction of particle propagation by 
\be
n_i\equiv \fr{P_i}{P}\qquad
\mbox{where}\qquad 
P^2\equiv \sum_{i=1}^3 P_i^2,
\lb{Pn_def_a}
\ee
so that
\be
\sum_{i=1}^3 n_i^2=1.
\lb{sum_ni_a}
\ee
We also introduce $n_0\equiv {P_0}/{P}$ 
(for the ultrarelativistic particles $g^{\mu\nu}n_\mu n_\nu=0$)
and $n^\mu\equiv g^{\mu\nu}n_\nu = P^\mu/P$.
The motivation for the noncovariant definitions\rf{Pn_def_a}
is to describe radiation transport
in terms of variables which are fully specified by
the ``dynamical'' quantity $f(\t,x^i,P_i)$
but are unmixed with the gauge-dependent and, 
in general, noncausally evolving metric~$g_{\mu\nu}$.

The energy-momentum tensor of radiation equals
\be
T^{\mu }_{\nu }
   =\int\fr{d^3P_i}{\rtg}\fr{P^\mu P_\nu}{P^0}\,f.
\lb{Tmunu_gen_a}
\ee
Substituting $P_\mu=n_\mu P$ and $P^\mu=n^\mu P$,
we can rewrite it as 
\be
T^{\mu }_{\nu}
   =\int\fr{\dOmn}{\rtg}\fr{n^\mu n_\nu}{n^0}\,I
\lb{Tmunu_int_a}
\ee
--a directional average of an expression which depends 
only on the local value of the metric and the quantity 
\be
I(x^\mu,n_i)\equiv \int_0^{\infty} P^3dP\,f(x^\mu,n_iP).
\lb{I_def_a}
\ee
The variable\rf{I_def_a}, to be called the {\it conformal intensity\/}
of radiation, inherits such useful properties of $f(x^\mu,P_i)$
as time-independence for superhorizon inhomogeneities of
non-interacting particles and 
reduction to the proper intensity,  $dE/(dVd^2\^{\bm n})$,
in the Minkowski limit.

The dynamics of~$I(x^\mu,n_i)$ for minimally coupled radiation in 
a known metric is given by a closed equation\ct{SB_tensors05}
which is straightforwardly derived by integrating 
the Boltzmann equation\rf{df_dot_free} over $P^3dP$ 
and remembering that for the ultrarelativistic particles 
$g^{\mu\nu}n_\mu n_\nu=0$:\footnote{
  In the last terms of eqs.\rf{dotI_gen_a} and\rf{doti_gen_a},
  the partial derivatives with respect to
  the components of $\nbrk{n_i=P_i/P}$, constrained by condition\rf{sum_ni_a}, 
  can be naturally defined as
$$
\fr{\pd }{\pd n_i}\equiv \sum_{\al=1,2}\lf(\fr{P\pd\mu_\al}{\pd P_i}\rt) 
    \fr{\pd}{\pd\mu_\al},
$$
  where $\mu_\al = (\mu_1,\mu_2)$ are any two  
  independent variables parameterizing~$\n$.
}
\be
n^\mu\fr{\pd I}{\pd x^\mu}= n^\mu n^\nu g_{\mu\nu,i}
     \lf(2n_iI-\fr12\fr{\pd I}{\pd n_i}\rt).
\lb{dotI_gen_a}
\ee
This equation confirms that for decoupled particles
which are perturbed only on superhorizon scales
(the gradients $\pd I/\pd x^i$ and $g_{\mu\nu,i}$ are negligible)
the conformal intensity freezes: $\nbrk{\dot I=0}$.

Now we can describe perturbation of the intensity 
by a variable 
\be
\di(x^\mu,n_i)\equiv \fr{I}{\bar I}-1,
\lb{di_def}
\ee
where $\bar I$ is a time-independent background value of~$I$.
As intended, the perturbation~$\di$ is time-independent
in full (nonlinear) general relativity for superhorizon perturbations
of uncoupled radiation. 
This variable also reduces to the perturbation of proper intensity 
in spatially homogeneous and isotropic, 
or Friedmann-Robertson-Walker (FRW), metric.
The full dynamical equation for $\di$ evolution
follows trivially
from eq.\rf{dotI_gen_a}:
\be
n^\mu\fr{\pd \di}{\pd x^\mu}= n^\mu n^\nu g_{\mu\nu,i}
     \lf[2n_i(1+\di)-\fr12\fr{\pd \di}{\pd n_i}\rt].
\lb{doti_gen_a}
\ee

The value of the introduced intensity perturbation $\di$ 
is generally gauge-dependent.
If spacetime is spatially homogeneous and isotropic 
then the FRW metric is naturally preferred 
as the metric manifesting the spacetime symmetries.
In this metric $\di$ is unambiguous
and coincides with the perturbation of proper intensity.
As shown next, the value of $\di$ is also unambiguous 
for superhorizon perturbations: it is the same in any of the gauges 
which provide a nonsingular description of the superhorizon evolution.
(More general arguments for the uniqueness
of ``physically adequate'' mapping of super- to subhorizon 
perturbations are given in Sec.~\ref{sec_form_mapping}.)

Under an infinitesimal change of coordinates 
$x^\mu\to \~x^\mu=x^\mu+\d x^\mu$ the  perturbation
$\di$ transforms as
\be
\~\di(\~x^\mu,\~n_i) = \di(x^\mu,n_i) - 4n_in_\mu \d x^\mu_{,i}, 
\lb{di_gtr}
\ee
where by eq.\rf{Pn_def_a}
\be
\~n_i = n_i - (\d_{ij}- n_in_j)n_\mu\d x^\mu_{,j}.
\lb{ni_gtr}
\ee
Suppose that in at least some gauges the {\it superhorizon\/} evolution
of cosmological perturbations appears {\it nonsingular\/} (regular). 
Namely, that in these gauges the perturbations of
all the dynamical and metric variables~$Q$ 
have a {\it moderate magnitude\/} $|\d|\equiv |\d Q/Q| \lesssim1$
{\it and are homogeneous\/} on the spatial scales
less than~$\l\sim |\d/\d_{,i}|\gg \H^{-1}$.
A transformation between two gauges either of which 
is regular, as defined here, 
has to be restricted as
\be
|\d x^\mu|\lesssim \H^{-1}\qquad {\rm and} \qquad
|\d x^\mu_{,i}|\lesssim \H^{-1}/\l.
\lb{reg_transf}
\ee
Given these restrictions and 
the result $\pd \di/\pd x^\mu = O(\l^{-1})$,
following from them and from eq.\rf{doti_gen_a}, 
the values of $n_i$ and $\di$ 
are unchanged by the gauge transformation\rfs{di_gtr}{ni_gtr}
in the superhorizon limit $\H^{-1}/\l\to 0$.\footnote{ 
  Most of the gauge-dependent measures of perturbations 
  vary in the $\H^{-1}/\l\to 0$ limit 
  even within the restricted class of regular gauges.
  For example, 
  the metric perturbation~$\d g_{\mu\nu}$ changes
  under a restricted gauge transformation $\~x^\mu=x^\mu+\d x^\mu$,
  with $|\d x^\mu_{,i}|\lesssim \H^{-1}/\l \to 0$,
  by $-\mathcal{L}_{\d x} g_{\mu\nu} =-(g_{\mu\l}\d^0_{\nu}
  -\d^0_{\mu}g_{\l\nu})  \d \dot x^\l-\dot{g}_{\mu\nu}\d x^0\not= 0$.
}

The equations given so far applied to the full 
general-relativistic dynamics.
We conclude this subsection by presenting
their linearized versions.
We consider the perturbed FRW metric
\be
g_{\mu\nu}=a^2(\t)\lf[\eta_{\mu\nu}+h_{\mu\nu}(x^\mu)\rt],
\lb{metric}
\ee
with $\eta_{\mu\nu}\equiv \diag(-1,1,1,1)$.
The linear in perturbation terms of eq.\rf{doti_gen_a} are
\be
\dot \di + n_i\di_{,i}  
  = 2n_i\^n^\mu \^n^\nu h_{\mu\nu,i},
\lb{dotD}
\ee
where $\^n^\mu \equiv (1,n_i)$.
The intensity perturbation~$\di$ 
determines linear perturbation of radiation coordinate density~$d$, 
introduced in the next subsection,
as [eq.\rf{di2d}]
\be
d = \fr34\,\la\di\ra_{\n},
\lb{di_0}
\ee
where $\la\ra_{\n}$ denotes $\int {\dOmn}/{4\pi}$.
Likewise,~$\di$ specifies normal bulk velocity of the radiation
\be
v_i\equiv  \fr{T^0_i}{\rho+p},
\lb{v_i_def}
\ee
where $\rho\equiv -T^0_0$ and $p\equiv \sum_i T^i_i/3$, \lb{rho_def}
as 
\be
v_i = \fr34\,\la n_i\di\ra_{\n},
\ee
which follows from eq.\rf{Tmunu_int_a}.

\subsection{Energy density}
\lb{sec_form_princip}

Next we consider a measure of perturbation of energy density  
for arbitrary species that, again,
is directly connected to local physical dynamics 
and is universally applicable to all scales.
While we lift all the restrictions on the nature of the species,
we limit our construction to the linear order of cosmological
perturbation theory.  For the discussion of densities
it is sufficient to address scalar perturbations
and disregard uncoupled vector and tensor modes.

Suppose that the studied, possibly self-interacting, species
(labeled by a subscript~$a$) 
do not couple non-gravitationally to
and are not created in significant amount
by gravitational decays of the remaining species.
Such species, to be called ``isolated'', can be assigned 
a covariantly conserved energy-momentum tensor~$T_a^{\mu\nu}$.
In an arbitrary gauge,
the covariant energy conservation $T_a^{0\mu }{}_{;\mu }=0$ 
in linear perturbation order gives
\be
\dot d_a + \pd_i v'^i_a =
    \fr{\dot\rho_a\d p_a-\dot p_a\d\rho_a}{(\rho_a+p_a)^2}.
\lb{dotd_any_gauge}
\ee
Here we introduced {\it coordinate number density\/} perturbation 
of the species
\be
d_a 
   \equiv \fr{\d\rho_a}{\rho_a+p_a} + 3D,
\lb{d_a_def}
\ee
where $D\equiv \fr16\,\d(\det g_{ij})/\det g_{ij} = {\textstyle \fr16\sum_i h_{ii}}$
is the spatial dilation of the metric\rf{metric},
and coordinate bulk velocity of the species
\be
v'^i_a \equiv  {-T^i_a{}_0 \ov \rho_a+p_a} = {v_a}_i-h_{0i}.
\lb{v_a_coo_def}
\ee
If the species~$a$ are classical particles
then $v'^i_a$ gives their momentum-averaged velocity 
measured by coordinate observers, moving along the lines~$x^i=\const$.

The justification for calling $d_a$ 
``coordinate number density perturbation'' is the following. 
Consider isolated species which
are described by an equation of state $p = p(\rho)$
and can be assigned a conserved number~$N$.
For them, by energy conservation, 
$\d\rho/(\rho+p)$ gives the relative perturbation  
$\d n_{{\rm pr}}/n_{{\rm pr}}$ of 
species' proper number density~$n_{{\rm pr}}=dN/dV\spr$.
Then $d_a$ equals the perturbation $\d n_{{\rm coo}}/n_{{\rm coo}}$ 
of species' {\it coordinate\/} number density~$n_{{\rm coo}}=dN/d^{3}\x$,
\lb{d_a_interp}
where the proper and coordinate infinitesimal volumes 
are related as $dV\spr=a^3(1+3D)d^{3}\x$.

The right-hand side of eq.\rf{dotd_any_gauge} is a gauge-invariant quantity,  
proportional to the so called ``nonadiabatic pressure'' 
$\d p_a-(\dot p_a/\dot\rho_a)\d\rho_a$.
It vanishes trivially 
whenever~$p_a$ is a unique function of~$\rho_a$.
Even if $p_a$ and $\rho_a$ are not functionally related,
for example, if $a$ is a field or a mixture of more basic 
interacting species,
the right-hand side of eq.\rf{dotd_any_gauge} 
vanishes whenever all the proper distributions describing the species~$a$ 
are unperturbed in certain coordinates.
Indeed, then $\dot\rho_a\d p_a-\dot p_a\d\rho_a$ 
vanishes in those coordinates and so, 
by its gauge invariance, vanishes in any coordinates.
We will call the superhorizon perturbations of the species~$a$ 
which appear unperturbed on a certain hypersurface {\it internally adiabatic\/}. 
We also note that the right-hand side of eq.\rf{dotd_any_gauge}
can be measured by a local
observer, provided he accepts a background equation
of state~$p_a = p_a(\rho_a)$
which defines the nature of the unperturbed species.

According to eq.\rf{dotd_any_gauge}, 
the density perturbation~$d_a$ is frozen
for superhorizon internally adiabatic perturbations of isolated 
species.  This superhorizon conservation can be easily understood
for particles with a conserved number, 
when  $d_a$ is literally the perturbation of 
coordinate density of this number.
As long as the chosen gauge is nonsingular
in the superhorizon limit $\H^{-1}/\l\to 0$, 
the particle number $n_{{\rm coo}}$
in a unit coordinate (but not proper!) volume
changes infinitesimally over a time interval $\D\t=\H^{-1}$, 
by $\D n_{{\rm coo}}/n_{{\rm coo}}= \D\t\,\pd_i v'^i_a=O(\H^{-1}/\l)$.
Hence, $d_a = \d n_{{\rm coo}}/n_{{\rm coo}}$ remains frozen.

Similarly to the perturbation~$\di$ of radiation intensity,
the coordinate density perturbation $d_a$ is generally gauge-dependent.
Yet, in the superhorizon limit its value in the regular gauges 
is also unambiguous.
This is evident from the $d_a$ transformation
\be
\~d_a = d_a - \d x^i_{,i}
\lb{d_a_trf}
\ee
under a change of coordinates $\~x^\mu=x^\mu+\d x^\mu$,
restricted for the regular gauges by conditions\rf{reg_transf}.

The traditional measures of density perturbation 
such as the perturbations of proper density or 3-curvature 
are, likewise, gauge-dependent.
But, unlike~$d_a$, those perturbations
vary in the superhorizon limit even among the regular gauges.
For example, a gauge transformation law
\be
\fr{\widetilde{\d\rho_a}}{\rho_a+p_a} = \fr{\d\rho_a}{\rho_a+p_a} +3\H \d x^0
\ee
shows that proper overdensities $\d\rho_a/(\rho_a+p_a)$ or $\d\rho_a/\rho_a$ 
remain gauge-dependent in the superhorizon limit 
among the regular gauges, restricted by 
conditions\rf{reg_transf}.

As noted in Sec.~\ref{sec_intro},
the values of the gauge-variant density, curvature, or other
perturbations in any fixed gauge
define certain ``gauge-invariant''
perturbations\ct{GerlachSengupta78,Bardeen80,KS84,BruniDunsbyEllis92}, 
such as Bardeen's $\eps_a$ or $\zeta$\ct{Bardeen80,zeta_orig}.
The value of $d_a$ in a fixed gauge also
defines a gauge-invariant variable.
In particular, $d_a$ of the Newtonian gauge (Sec.~\ref{sec_newt_dyn})
corresponds to the gauge-invariant expression\rf{d_a_ginv}.

We can simply relate 
the perturbation of the coordinate particle number density~$d$
with the discussed earlier 
perturbation of phase-space distribution~$d f(x^\mu,P_i)$ 
for classical point particles
and with perturbation of intensity~$\di$
for ultrarelativistic particles.
As follows from 
eqs.\rf{d_a_def} and\rf{Tmunu_gen_a},
in linear order,
\be
d=\fr{\int d^3P_i\,E(P)\,d f}{a^4(\rho+p)},
\lb{df2d}
\ee
where $E(P)\equiv (P^2+m^2a^2)^{1/2}$.
For ultrarelativistic particles
this equation and eqs.\rf{I_def_a} and\rf{di_def} give
\be
d = \fr34\,\la\di\ra_{\n}.
\lb{di2d}
\ee

Since the variable~$d_a$ is constant
in a regular gauge for superhorizon internally adiabatic perturbations 
of isolated species, any change of~$d_a$
can be linked to an objective dynamical cause 
within the local Hubble volume.
We show next that the variable~$d_a$ 
remains useful on all scales
and that the suggested by it relation between density 
perturbations on super- and subhorizon scales 
is uniquely preferable on physical grounds.

\subsection{Uniqueness of superhorizon values}
\lb{sec_form_mapping}

Various measures of scalar perturbations
that freeze beyond the horizon,
for example, the uniform-density or comoving
curvatures $\zeta$\ct{Bardeen80,zeta_orig} 
or $\mathcal{R}$\ct{Bardeen80,KS84,Lyth84}
or the gradients of Refs.\ct{RigopoulosShellard03,Langlois05},
differ in their evolution on the scales comparable to 
the horizon and smaller.
Not all of these measures of perturbations
remain meaningful in the subhorizon Minkowski limit.
Those that do,
\eg\ the coordinate density perturbations~$d_a$
in various regular gauges,
may still differ from each other 
on scales comparable to the Hubble scale.

We now prove that,
nevertheless, 
{\it the value of a superhorizon density perturbation is the same\/}
in terms of any measure which
\begin{enumerate}
\item[I$'$.]\lb{modif_conds}
Freezes for superhorizon internally adiabatic 
perturbations of uncoupled (isolated) species,
\end{enumerate}
and additionally,  
\begin{enumerate}
\item[II$'$.]
Gives the perturbation of species' number density, 
$\d n_a/n_a \equiv \d\rho_a/(\rho_a+p_a)$,
in any spatially homogeneous and isotropic (FRW) metric.
\end{enumerate}
Condition~II$'$ is a formal statement
of our desire for a measure of density perturbation
to be such on all scales 
whenever the concept of density perturbation is unambiguous
(\cf requirement~II in Sec.~\ref{sec_intro}).
We match the measure to $\d\rho_a/(\rho_a+p_a)$,
rather than, \eg, to $\d\rho_a/\rho_a$,
to be consistent with condition~I$'$, 
which then applies automatically to an FRW metric.
Finally, if the dynamics of the measure is fully specified
by the physics within the local Hubble volume 
(requirement~I in Sec.~\ref{sec_intro}),
then the local dynamics of superhorizon perturbations 
should be the same whether the metric is
globally FRW or not.
Hence, we impose condition~I$'$
for the general metric.

The proof is applicable to arbitrary isolated species~$a$. 
To be specific, we consider a decelerating universe,
where perturbations evolve from super- to subhorizon scales. 
We {\it imagine\/} a cosmological scenario 
in which the initially perturbed
geometry becomes homogeneous and isotropic
while the studied scales~$\l$ are  superhorizon
($\nbrk{\l/\H^{-1}\gg1}$)
and remains  homogeneous and isotropic 
ever since.
We suppose that the chosen gauge provides a regular 
description of superhorizon perturbations
and leads to the FRW metric in the unperturbed geometry.
Let $d_a$ and $d'_a$ be two arbitrary measures that
both are frozen on superhorizon scales and
equal to $\d\rho_a/(\rho_a+p_a)$ in the FRW metric.
Since in our imaginary scenario 
the metric becomes FRW on superhorizon scales,
$d_a=d'_a$ at all times.
In particular, $d_a=d'_a$ beyond the horizon.
We now note that if the measures describe 
an instantaneous state of the perturbations
then the equality $d_a=d'_a$
retains its validity on superhorizon scales
regardless of the subsequent perturbation evolution,
\ie, regardless of the chosen cosmological scenario.

To conclude the proof, we demonstrate that 
the invoked imaginary scenario is physically realizable,
however unnatural this realization may be.
Given arbitrary superhorizon perturbations
of the existing cosmological species and the metric,
we can force them to evolve to
a homogeneous geometry by adding
new minimally coupled species,~$X$, which first contribute 
infinitesimally to the overall energy-momentum density.
Let these imaginary species obey an equation of state
$w_X\equiv p_X/\rho_X = -1/3$ until they become the only
dominant component.
(The choice $w=-1/3$ 
corresponds to the borderline
between deceleration and acceleration,
hence, time-independent $\l/\H^{-1}$.)
We set the coordinate number density perturbation~$d_X$ 
of the species~$X$ to vanish on superhorizon scales, 
\eg, by our choice of the initial distribution of these species.
(This requirement is not dynamically natural.
Yet, due to superhorizon conservation of~$d_X$,
it is self-consistent and in principle realizable.)
Then, once the unperturbed species~$X$ dominate,
they generate a homogeneous and isotropic geometry.
The subsequent cosmological expansion may be
designed to decelerate again, \eg, 
by taking $w_X$ at any $\l/\H^{-1}$ equal $w$
at the same $\l/\H^{-1}$
in the original model, without the species~$X$.
Then in the modified model the fictitious species~$X$ remain dominant
and the geometry unperturbed, as desired.
We stress that this scenario and the species~$X$
are introduced only as a gedanken experiment.
They are not expected to exist in nature.\footnote{
  Since the appearance of the first astro-ph version of this paper 
  (astro-ph/0405157)
  the suggested scenario has been considered 
  by several authors\ct{Bartolo_TS05,Sloth_TS05}
  for damping scalar perturbations,
  hence, enhancing the ratio of tensor to scalar modes after inflation.
  Ref.\ct{LindeMukh_TS05}, however, points out that
  the required superhorizon perturbations
  of the species~$X$ are highly unnatural to be produced dynamically
  in the real world.
}

The above arguments do not imply that the measures $d_a$ and $d'_a$ 
are identical.  For example, the coordinate number density
perturbations\rf{d_a_def} differ in different gauges,
\cf eq.\rf{d_a_trf}.
The identical are their values 
for superhorizon perturbations
and, trivially, for unperturbed geometry,
when they all give $\d\rho_a/(\rho_a+p_a)$.

\section{Dynamics in the Newtonian Gauge}
\lb{sec_newt_dyn}

Any convenient gauge can be used to study
cosmological evolution.
Nevertheless,
of the most popular gauges, such as the synchronous, Newtonian,
uniform density, comoving, or spatially flat,
only the Newtonian gauge reduces to Newtonian gravity on subhorizon scales. 
Because of this helpful feature, 
during our further discussion
of scalar perturbations we impose the Newtonian gauge conditions,
namely, require zero shift and shear of the metric: 
$g_{i0}\equiv0$ and $g_{ij}\propto\d_{ij}$.
For simplicity, we assume a spatially flat background geometry.
Then the perturbed metric reads
\be
ds^2=a^2(\tau )\,[-(1+2\Phi )d\tau ^{2}+(1-2\Psi )d\x^{2}]\,.
\lb{Newt_gauge_def}
\ee
(It is not a priori obvious that 
the Newtonian gauge provides a nonsingular description
of superhorizon scales.  We will return to this subtlety 
at the end of the section.)

\subsection{Densities}
\lb{sec_dens}

In the Newtonian gauge
the coordinate number density perturbation\rf{d_a_def} is 
\be
d_a = \fr{\d\rho_a}{\rho_a+p_a} - 3\Psi,
\lb{d_a_Newt}
\ee
and the coordinate and normal bulk velocities of species 
coincide: $v'^i_a=v_{i\,a}$ [eq.\rf{v_a_coo_def}].

The Newtonian-gauge variable\rf{d_a_Newt}
can be written in a gauge-invariant form
\be
d^{(\rm Newt)}_a= \fr{\d\rho_a}{\rho_a+p_a}+3D+\nabla^2\epsilon,
\lb{d_a_ginv}
\ee
where $D$\ and $\epsilon$ parameterize the scalar perturbations of
the spatial part of the metric as
\be
\d g_{ij}\equiv a^{2}[2D\,\d_{ij}-2(\nabla_{\!i}\nabla_{\!j}
            -\frac{1}{3}\d_{ij}\nabla^{2})\,\epsilon ].
\lb{g_ij_param}
\ee  
Evaluation of the gauge-invariant combination
on the right-hand side of eq.\rf{d_a_ginv} 
in other gauges shows that:
\begin{itemize}
\item[a)]
$d^{(\rm Newt)}_a=3\zeta_a$, 
where $\zeta_a$\ct{zeta_a} is the (reduced) curvature
$\zeta_a=D+\fr13\Nb^2\epsilon=-\fr14\,a^2(\Nb^{-2}){\,}^{(3)}\!R$ 
on the hypersurfaces of uniform density of species~$a$ 
(on which $\rho_a(\x)\equiv\const$);
\item[b)]
$d^{(\rm Newt)}_a$ equals
$\d n_{a\,{\rm pr}}/n_{a\,{\rm pr}}=\d\rho_a/(\rho_a+p_a)$ 
in the spatially flat gauge ($\d g_{ij}\equiv 0$);
\item[c)]
$d^{(\rm Newt)}_a$ coincides with $d_a\equiv \d n_{a\,{\rm coo}}/n_{a\,{\rm coo}}$
of eq.\rf{d_a_def} in any gauge with zero shear ($\epsilon\equiv 0$),
regardless of the second gauge-fixing condition.
\end{itemize}

The evolution of density perturbations~$d_a$
in any gauge, including the Newtonian, 
is described by eq.\rf{dotd_any_gauge}.
To complete the formalism, we provide the Newtonian gauge equations 
for the evolution of species' velocity,
pressure, etc.\ and for the metric perturbations.

\subsection{Velocities}
\lb{sec_veloc}

For scalar perturbations, we introduce a velocity potential~$u_a$
such that 
\be
{v_a}_i \equiv -\Nbi u_a.
\lb{vp_def}
\ee
Covariant momentum conservation $T_a^{i\mu}{}_{;\mu}=0$ 
for isolated species gives
\be
\dot u_a = -\, \H u_a + \Phi
    + \fr{\d p_a-\dot p_a u_a}{\rho_a+p_a}+{\textstyle \fr23}\Nb^2\s_a,
\lb{dot_u}
\ee
where a scalar potential~$\s_a$ 
parameterizes the species' anisotropic stress 
$\Sigma_a^i{}_j\equiv T_a^i{}_j-\fr13\d^i_jT_a^k{}_k$ as \lb{Sigma_def}
\be
\fr{\Sigma_a^i{}_j}{\rho_a+p_a} \equiv 
(\Nbi\Nbj-{\textstyle \fr13}\d_{ij}\Nb^2)\s_a.
\lb{sigma_def}
\ee

The pressure perturbation $\d p_a$
and anisotropic stress potential~$\sigma_a$ are 
determined by the internal dynamics of the species.
In Appendix~\ref{sec_dyn} we give the complete 
Newtonian-gauge equations for the linear evolution 
of scalar perturbations
of typical cosmological species: CDM,
decoupled neutrinos, partially polarized photons, baryons,
and a classical scalar field.
In particular, for the following discussion,
we note that the perturbation of intensity 
of decoupled radiation 
(neutrinos or CMB photons after their last scattering)
evolves in the Newtonian gauge as
\be
\dot \di + n_i\Nbi \di  
  = -4 n_i\Nbi(\Phi+\Psi),
\lb{dot_di_Newt}
\ee
as follows from eq.\rf{dotD}.

\subsection{Metric}
\lb{sec_metric}

The Newtonian gravitational potentials $\Phi$ and $\Psi$
do not represent additional dynamical degrees of freedom but
are fully determined by matter perturbations.
To write the corresponding equations\ct{BS}, 
we introduce the reduced background enthalpy density 
\be
\g \equiv \nbrk{4\pi Ga^2(\rho+p)}=
{\textstyle \fr{3(1+w)}2}\,\H^2
\lb{gamma_def}
\ee
(with $\rho=\sum \rho_a$, $p=\sum p_a$ and $w\equiv p/\rho$),
species enthalpy abundances
$x_a\equiv({\rho_a+p_a})/({\rho+p})$, 
and enthalpy-averaged perturbations
\be
d \equiv \fr{\d\rho}{\rho+p} + 3D = \sum x_a d_a, 
\ee
$\nbrk{u=\sum x_a u_a}$
and $\nbrk{\s=\sum x_a \s_a}$
[for the latter two, $\nbrk{v_i\equiv T^0_i/(\rho+p) = -\Nbi u}$
and $\nbrk{(T^i_j-\fr13\d^i_jT^k_k)/(\rho+p)}=\nbrk{(\Nbi\Nbj-\fr13\d_{ij}\Nb^2)\s}$].
Then the linearized Einstein 
equations\ct{Mukh_Rept,MaBert95} become
\be
\Nb^2\Psi-3\g\Psi&=&\g\lf(d+3\H u\rt),
\lb{Psi_eq}\\
{\textstyle \fr12}(\Psi-\Phi)&=&\g\s.
\lb{Phi_eq}
\ee

In real space eq.\rf{Psi_eq} is solved by
\be
\Psi(\r)=-\,\fr{\g}{4\pi}\int d^3\r'\,\fr{ e^{-r'\!/r_G}}{r'}\lf.\lf(d+3\H u\rt)\rt|_{\r+\r'},
\lb{Psi_real_sol}
\ee
with $r_G\equiv 1/\sqrt{3\g}=\fr13(\fr2{1+w})^{1/2}\H^{-1}$. 
For subhorizon perturbations 
the term $\H u$ is negligible and $e^{-r'\!/r_G}\approx 1$.
Then the induced potential $\Psi(\r)$ obeys the usual $1/r$ Newton's law
(and so does $\Phi\approx\Psi$).
However, on large, compared to $r_G$, scales $\Psi$ is fully determined 
by the {\it local\/} coordinate overdensities and velocity potentials
of the matter species: 
\be \qquad\qquad\qquad
\Psi\simeq-{d\ov3}-\H u \qquad (\mbox{for}~r\gg r_G),
\lb{Psi_superhor}
\ee
as follows from either eq.\rf{Psi_real_sol} or eq.\rf{Psi_eq}.
The corrections to relation\rf{Psi_superhor} 
from matter inhomogeneities beyond the distance~$r_G$
are suppressed exponentially.

The presented dynamical equations for
the coordinate measures of density and intensity perturbations
have three important advantages over the traditional Newtonian-gauge formalisms
in terms of the proper measures, \eg\ct{MaBert95,Dod_book}.
First of all, by Sec.~\ref{sec_form},
the evolution of the coordinate, rather than proper, 
measures of perturbation is
determined entirely by the physics within 
the local Hubble volume.

The second, related to the first, 
improvement of the formalism is the explicit Cauchy structure
of the new dynamical equations in the Newtonian gauge.
Indeed, the traditional equations for the rate of change of 
proper density or intensity perturbations  
inevitably contain the time derivative 
of~$\Psi$\ct{MaBert95,HuSugAnalyt95,CMBFAST96,ZalSel_allsky97,Dod_book}.
Yet, by the generalized Poisson equation [\cf eq.\rf{Poiss_tradit}],
$\dot\Psi$ itself depends on the rate 
of change of the density and velocity perturbations of all the species.
The terms $\dot\Psi$ or $\dot\Phi$ do not appear 
in the proposed equations.

And the third benefit of quantifying
density perturbation by $d=\d n_{{\rm coo}}/n_{{\rm coo}}$
is a nonsingular generalized Poisson equation for~$\Psi$, eq.\rf{Psi_eq}.
In terms of the traditional variables,
\be
\Nb^2\Psi = \g\lf(\fr{\d\rho}{\rho+p}+3\H u\rt).
\lb{Poiss_tradit}
\ee
As $\t\to0$, the factor $\g$, eq.\rf{gamma_def}, 
diverges as $\H^2\sim 1/\t^2$.
For the typical, \eg\ inflationary, initial conditions
this divergence on superhorizon scales is counterbalanced by 
a cancelation $\fr{\d\rho}{\rho+p}+3\H u = O(\t^2)$.
The seemingly singular right-hand side of 
traditional eq.\rf{Poiss_tradit}
may misguide an analytical analysis
as well as cause significant numerical errors and instabilities
due to incomplete cancelation of the singular contributions.
The regularity of $\Psi$ in the superhorizon limit,
on the other hand, 
is natural from eq.\rf{Psi_eq} for any conserved overdensities~$d_a$.
There is generally no large cancelation in the superhorizon limit of
eq.\rf{Psi_eq}.

We now return to the question of whether
the Newtonian gauge provides 
a nonsingular description of superhorizon perturbations.
As proved in Ref.\ct{Weinb_adiab1}, for any reasonable cosmology
there exist perturbation solutions
for which species' overdensities, the product $\H u$,
and gravitational potentials in the Newtonian gauge  
remain finite in the superhorizon limit.
In addition, eq.\rf{dot_u} admits singular solutions
with $u\sim 1/a$.
One might discard the singular modes on the grounds
that if they are comparable to the regular ones
when the perturbations exit the horizon
then the singular modes become negligible soon after the exit.
However, the solutions which appear singular
in the Newtonian gauge could be produced 
by a physical mechanism 
whose description is singular in the Newtonian
but regular in a different gauge.
Then on superhorizon scales
in the Newtonian gauge 
decaying singular modes
would appear to dominate frozen regular modes.

This question cannot be resolved without taking 
into account dynamical equations. 
For example, we can imagine regular perturbations
in metric whose shear
does not vanish ($\Nb^2\eps\not\to 0$) in the superhorizon limit.
The transformation of metric shear 
\be
\Nb^2\~\eps=\Nb^2\eps + \d x^i_{,i}
\ee
under a scalar gauge transformation
shows that this metric requires  
a singular transformation to 
the Newtonian gauge, in which $\eps^{\rm (Newt)}\equiv0$.
The corresponding Newtonian perturbations may then turn out singular.
Fortunately, we know that, at least, 
slow-roll inflationary scenarios lead to 
superhorizon perturbations which are nonsingular
in the Newtonian gauge\ct{Mukh_Rept}.
Therefore, given the convenience of the Newtonian gauge,
we use it for our analysis 
of post-inflationary cosmological evolution.

\section{Overview of Perturbation Evolution}
\lb{sec_evol}

\subsection{Superhorizon evolution}
\lb{sec_evol_superhor}

We can identify three popular descriptions
of perturbation evolution on superhorizon scales.
Following tradition, the general superhorizon perturbation
is often viewed as a superposition of adiabatic and isocurvature modes,
\eg\ct{Kolb_textbook}. 
Adiabatic (also ``curvature'' or ``isentropic'') perturbations can be defined
as perturbations for which the internal (proper) characteristics of 
{\it all\/} the matter species are uniform
on some spatial slice.
On this slice the metric remains perturbed.
Isocurvature perturbations were idealistically considered
to be the perturbations of the relative abundances of species  
while the total energy density was assumed
unperturbed  and the geometry homogeneous and isotropic.
It has been, however, realized for two decades\ct{Linde_curv84,Linde_curv85,
Kofman_curv85,Mollerach_curvaton89,LindeMukh_curv96} 
that the decomposition into adiabatic and isocurvature modes
generally is not invariant under cosmological evolution.
Although an adiabatic perturbation preserves its nature
on superhorizon scales\ct{Weinb_adiab1,BS,Weinb_adiab2},
a perturbation which appears isocurvature at a particular time,
generally fails to be such in just a single e-folding of Hubble expansion.
Despite this serious inconsistency, the picture of ``isocurvature modes''  
continues to be influential, at least, in terminology.

A different view of superhorizon evolution
is given by the ``separate universe'' 
picture, \eg\ct{Starobinsky_sepuniv86,SasakiStewart_sepuniv95,LiddleLyth_book}.
In it, each Hubble volume is considered as an independent FRW universe,
characterized by its own expansion rate and curvature.
This description is self-consistent,
and it explicitly exhibits the locality of superhorizon evolution. 
Nevertheless, it requires to keep track of ``how far'' 
each region has evolved in a global frame.
The procedure for the latter is well defined
but its outcome is not intuitive
when the perturbations are not adiabatic.

More recently, Wand et al.\ct{zeta_a} noted that  
superhorizon evolution of several uncoupled fluids 
becomes trivial 
if inhomogeneity of a given fluid~$a$ 
is described by the perturbation of spatial curvature 
on the hypersurfaces of constant~$\rho_a$.
Neglecting gravitational decays of species\ct{LindeMukh_curvaton05},
these curvature perturbations,~$\zeta_a$, are individually conserved
for uncoupled fluids
even if the overall perturbation is not adiabatic.
This approach lead to significant progress, 
\eg\ct{Lyth:2001nq,Lyth:2002my},
in quantitative analysis of 
the curvaton-type scenarios\ct{Linde_curv84,Linde_curv85,
Kofman_curv85,Mollerach_curvaton89,LindeMukh_curv96}.

The formalism presented in Secs.~\ref{sec_form} and~\ref{sec_newt_dyn} 
further facilitates the description of linear superhorizon evolution 
of multiple species as the common expansion 
of gravitationally decoupled inhomogeneities in the individual species.
It shows that if the inhomogeneities are quantified
by the perturbations of the coordinate number densities
then their superhorizon freezing
is explicit in any regular gauge.
The conserved curvatures~$\zeta_a$\ct{zeta_a}
are simply related to the coordinate density perturbations~$d_a$ 
in the Newtonian gauge, as $\zeta_a=\fr13\,d^{(\rm Newt)}_a$
[comment a) after eq.\rf{g_ij_param}].
The description in terms of~$d_a$
does not require to specify the conserved perturbations
of each fluid~$a$ on a different set of hypersurfaces.
This description retains its usefulness on all scales,
including subhorizon.

If the superhorizon cosmological perturbations
are adiabatic, as predicted by single-field inflation
and favored by observations\ct{WMAPPeiris,IsocurvCrotty03}, 
then the number density perturbations~$d_a$ of all the species are equal.
This can be seen by evaluating the right-hand side of eq.\rf{d_a_ginv}
in the gauge in which all $\d\rho_a$ vanish.
The common value of $d_a$ then equals three times the perturbation 
of the uniform-density slice curvature~$D$,
known as the Bardeen curvature~$\zeta$\ct{Bardeen80,zeta_orig}:
$d_a=3\zeta$.
In other words, on superhorizon scales
an arbitrary collection of adiabatically perturbed species 
is equivalent to a single fluid,
having an equation of state $p(\rho)=\bar p(\bar\rho)$ 
and frozen coordinate number density perturbation $d=3\zeta$.

Even if the overall cosmological perturbations are not adiabatic,
any species which do not interact with the other species
and are perturbed internally adiabatically
(\ie, are unperturbed on a certain spatial slice,
on which the other species may be perturbed) 
can be regarded as a single fluid, 
``same'' in various regions beyond causal contact.
Its density perturbation~$d_a$ is also frozen
(Sec.~\ref{sec_form_princip}).

The bulk peculiar velocities of species change with time
even on superhorizon scales. 
In general, the evolution of velocities differs for various species.
However, if superhorizon perturbations are adiabatic 
then the last two terms in eq.\rf{dot_u} vanish.
(These terms are gauge-invariant and 
they apparently vanish for adiabatic perturbations 
in the uniform density gauge.) 
Then for a Fourier perturbation mode with a wavevector~$k$  
\be
\fr{\pd}{\pd\t}(au_a) = a[\Phi + O(k^2/\H^2)].
\lb{v_superhor}
\ee
Thus for adiabatic conditions the superhorizon acceleration 
of all the species is equal.

\subsection{Horizon entry}
\lb{sec_evol_entry}

When the perturbation scale becomes comparable to the Hubble horizon
(the perturbations ``enter the horizon''),
the contribution of velocity divergence 
to $\dot d_a$, eq.\rf{dotd_any_gauge}, becomes appreciable.
The evolution of the velocities
is affected by metric inhomogeneities
and the latter are sourced by all the species 
which contribute to $\d T^{\mu\nu}$.
Thus, during horizon entry
all such species influence the evolution of
all the density and velocity perturbations.

After the entry,
the perturbations of given species may or may not
continue affecting other species gravitationally.
The criterion for perturbations to continue being
gravitationally relevant is evident
from eqs.\rfs{Psi_eq}{Phi_eq}.
On subhorizon scales these equations reduce to 
the Poisson equation
$k^2\Phi\simeq k^2\Psi\simeq
 - 4\pi Ga^2(\rho+p)\,d
=- 4\pi Ga^2 \sum_a(\rho_a+p_a)d_a$.
If for some species after the horizon entry
$a^2(\rho_a+p_a)d_a$ decays,
such as for relativistic species,
then the species contribute negligibly
to the gravitational potential.
The gravitational impact of their perturbations
can be ignored.
On the other hand, if 
$a^2(\rho_a+p_a)d_a\sim a^2\d\rho_a$ remains non-vanishing,
such as for CDM during the matter era,
then these species source metric perturbations
and maintain their influence on the evolution of all cosmological species.

\section{Features in the Angular Spectrum of the CMB}
\lb{sec_CMBfeatures}

\subsection{Preliminaries}
\lb{sec_CMB_prelim}

Prior to decoupling of photons from baryons at $z\sim 1100$,
Fourier modes of photon overdensity 
evolve according to wave equation\rf{dot_g} in Appendix~\ref{sec_dyn}.
We find it useful to rewrite this equation as
\be
\textstyle
\ddot d\sg + \fr{R_b \H}{1+R_b}\,\dot d\sg + c_s^2k^2(d\sg-B) = 0,
\lb{dot_gk}
\ee
where $R_b=3\rho_b/(4\rho_{\g})$, \lb{R_b_def}
$c_s^2=[3(1+R_b)]^{-1}$ gives the sound speed in the photon-baryon plasma,
\be
B(\t,k)\equiv\nbrk{-3(\Phi+\Psi+R_b\Phi)},
\lb{B_def}
\ee
and we continue to use the Newtonian gauge\rf{Newt_gauge_def}.
According to eq.\rf{dot_gk},
in each mode the density perturbation~$d\sg$
oscillates about its varying equilibrium value~$B(\t,k)$.

The presently observed CMB perturbations
are given by a line-of-sight solution\ct{CMBFAST96} of 
the radiation transport equation\rf{dot_di_Newt}
complemented by the Thomson collision 
terms\ct{Kaiser83,BondEfstathiou84,MaBert95}, eq.\rf{dot_di_g}.  
Denoting the collision terms by $C_T$, we have
\be
\dot \di\sg + n_i\Nbi \di\sg 
  = -4 n_i\Nbi(\Phi+\Psi) + C_T.
\lb{dot_dig_C}
\ee
Once the evolution of the dynamical variables
$d\sg$, $d\snu$, $d_c$, $u\sg$, etc.\ and the induced
gravitational potentials $\Phi$ and~$\Psi$ are  determined,
the line-of-sight integration of the radiation transport equation\rf{dot_dig_C}
becomes more straightforward if the spatial derivatives
of the potentials in eq.\rf{dot_dig_C} are traded for their time derivatives as
\be
\dot\di\eff + n_i\Nbi\di\eff = 4(\dot\Phi+\dot\Psi) + C_T,
\lb{dot_dieff_C}
\ee
where 
\be
\di\eff\equiv \di\sg + 4(\Phi+\Psi).
\lb{dieff_def}
\ee

The variables $\di\sg(\n)$ and $\di\eff(\n)$
have identical multipoles for all $\ell\ge1$.
However, their monopoles differ:
$\langle\di\sg\rangle_{\n}=\fr43d\sg$, eq.\rf{di_0},
while $\langle\di\eff\rangle_{\n}=4\Teff$,
where the effective temperature perturbation 
$\Teff$\ct{KodamaSasaki86,HuSug_ISW94,HuSug_toward94} 
in our variables reads
\be
\Teff \equiv \fr13\,d\sg+\Phi+\Psi.
\lb{Teff_def}
\ee
(In the traditional approach\ct{HuSug_ISW94,HuSug_toward94}, 
$\Teff= {\d T^{(\rm Newt)}}\!/{T}+\Phi$.)
Physically, $\Teff$ quantifies the present temperature perturbation 
of photon radiation which decoupled from a stationary thermal region 
with photon overdensity~$d\sg$ and 
potentials $\Phi$ and~$\Psi$,
provided the potentials subsequently decayed adiabatically slowly
over a time interval $\D\t\gg \l \equiv k^{-1}$.
Indeed, under the described conditions
$\di\eff$ is conserved along the line of sight by eq.\rf{dot_dieff_C}.\footnote{
  The variables $\di\eff$ and $\Teff$
  are useful for constructing the CMB line-of-sight transfer 
  functions\ct{CMBFAST96} from {\it known\/} solutions for perturbation modes.
  Yet, these variables 
  are not convenient as primary dynamical variables
  and should be avoided when studying
  perturbations' gravitational interaction. 
  The evolution equations for $\di\eff$ and $\Teff$
  do not have an explicit Cauchy form 
  and their dynamics does not
  satisfy condition~I of Sec.~\ref{sec_intro}.
}

The appearance of the term $\Phi+\Psi$,
rather than the traditional~$\Phi$, in eq.\rf{Teff_def}
should not be surprising. 
The same sum $\Phi+\Psi$ is also responsible 
for the integrated Sachs-Wolfe\ct{SachsWolfe67} (ISW) 
corrections to the intensity in the variable potential and 
for the deflection of photons by gravitational lensing.\footnote{
  Of the numerous existing formalisms for CMB dynamics in the Newtonian gauge, 
  I am aware of only one\ct{Durrer_pert93,Durrer_CMB01}, by R.~Durrer, 
  in that the impact of metric perturbations on radiation temperature 
  is described by $\Phi+\Psi$, as opposed to~$\Phi$.
  Durrer's variable
  for linear perturbation of radiation intensity 
  coincides with the linear limit  
  of our intensity perturbation~$\di$,
  constructed in Sec.~\ref{sec_form_intens} for full theory.
  Yet, unlike the considered~$d_a$,
  her density perturbations~$D_a=(1+w_a)d^{(\rm Newt)}_a$
  {\it do not freeze\/} for superhorizon perturbations.
}

The full expression for the CMB temperature perturbation
observed at the present time~$\t_0$ reads:
\be
\fr{\D T}{T}=\int_0^{\t_0}\!d\t&&\!\!\!\!\!\lf[\,
             \dot g \lf(\Teff + n_iv_b^i + 
             \fr14n_in_j\Pip^{ij}\rt)\rt.\!+ 
\nn\\&&
+\lf.g\lf(\dot\Phi+\dot\Psi\rt)\rt]_{\t,\,\r(\t)}~.
\lb{los_zeta}
\ee
It follows from the line-of-sight integration\ct{CMBFAST96} 
of eq.\rf{dot_dieff_C}
with the scattering terms of eq.\rf{dot_di_g}.
Here, $g(\t)$ is the probability for a photon 
emitted at time~$\t$
to reach the observer unscattered, eq.\rf{visfun_def},
and for the scalar perturbations 
$\Pip^{ij}=(\Nbi\Nbj-\fr13\d_{ij}\Nb^2)\pip$,
where $\pip$ for Thomson scattering is given by eq.\rf{q_def}.
The perturbations $\Teff$, $v_b$, and $\pip$
are evaluated along the line of sight
$\r(\t)=(\t-\t_0)\n$,
taking that the direction of $\D T$ observation is $-\n$
and at the observer's position $\r\equiv 0$.
The complete formulas for 
the angular power spectra $C_l$ of CMB temperature and polarization 
induced by the scalar perturbations
are summarized in Appendix~\ref{sec_Cl}.

We now proceed to connecting 
features in the angular spectrum of CMB temperature anisotropy
to the internal nature of cosmological species.
In Sec.~\ref{sec_rad} we address the scales
which enter the horizon during radiation domination,
and in  Sec.~\ref{sec_mat} the larger scales
which enter during matter and dark energy domination.

\subsection{Radiation era}
\lb{sec_rad}

\subsubsection{Radiation driving?}
\lb{sec_rad_driv}

It is often stated, \eg\ct{HuSugSmall95,HuNature,HuFukZalTeg00,HuDodelson02,Dod_book,Mukhanov_book},  
that the CMB perturbations on the scales of a degree
and smaller ($\ell\gtrsim 200$) are significantly enhanced
during the horizon entry in the radiation era. 
In a model without neutrinos,
the claimed enhancement (``radiation driving'')
of CMB temperature perturbations is by a factor of 3, 
translating into a factor of 9 enhancement 
of the temperature autocorrelation spectrum.
This conclusion is based on the apparent 
enhancements of both $\Teff$ and $\d T/T$ in the Newtonian gauge
during the entry
(the solid green and dashed curves in Fig.~\ref{fig_rad_driving}).
The enhancements are ascribed to a ``resonant boost''
of CMB perturbations by their own gravitational potential, whose decay 
soon after the horizon entry is said to be
``timed to leave the photon fluid maximally 
compressed''\ct{HuDodelson02}.

\begin{figure}[t]
\includegraphics[width=4.25cm]{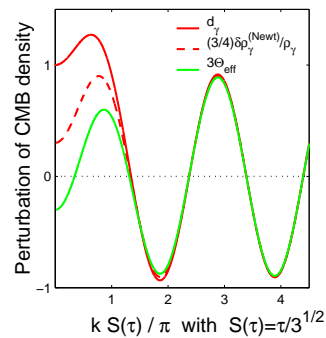}
\caption{
The evolution of a Fourier mode 
of CMB density perturbation in the radiation era
in terms of different measures:
Coordinate number density 
perturbation~$d\sg=\d n_{\g\,{\rm coo}}/n_{\g\,{\rm coo}}$ (solid),
proper density perturbation~$\d\rho\sg/\rho\sg$  
in the Newtonian gauge (dashed),
and effective temperature 
perturbation~$\Teff$, eq.\rf{Teff_def}, (green solid),
all normalized to coincide on small scales.
}
\lb{fig_rad_driving}
\end{figure}

However, in the concordance cosmological model as much as $41\%$ of the 
energy density in the radiation era is carried by 
decoupled neutrinos.  The evolution 
of free-streaming neutrinos and their contribution to metric inhomogeneities
differs qualitatively\ct{BS} from the acoustic dynamics of the photon fluid.  
If without neutrinos the CMB perturbations were resonantly
boosted by metric inhomogeneities, 
neutrinos should be expected to significantly untune the ``resonance''.
In reality, neutrinos decrease subhorizon $\d T/T$ 
by as little as~$10\%$, as compared to subhorizon $\d T/T$ in a model with
identical inflation-generated primordial perturbations but no neutrinos.
Moreover, the addition of noticeable density of early tracking 
quintessence\ct{RatraPeebles88,FerreiraJoyce97,
ZlatevStein_tracking98,OttExpQuint01}, 
whose perturbations evolve very dissimilar 
to the acoustic CMB perturbations
(dashed vs solid lines in Fig.~\ref{fig_evol_rad}), 
not only fails to destroy the ``resonance'' 
but instead enhances subhorizon~$\d T/T$ (Fig.~\ref{fig_impact_rad}\,a).

We argue that the apparent radiation driving
of photon temperature in the Newtonian gauge 
is a {\it gauge artifact\/}. 
First, the driving is absent when CMB density perturbations are 
quantified by the measure~$d\sg$, satisfying conditions~I and~II.
Indeed, in a photon-dominated universe
the overdensity $d\sg$ in the Newtonian gauge evolves as\ct{BS}
\be
d\sg = d\i\lf(\fr{2\sin\f}{\f}-\cos\f\rt),
\quad  
\f\equiv k\t/\sqrt3,
\lb{d_rad}
\ee 
as follows from eqs.\rf{dot_gdu} and\rf{Psi_eq}.
Hence, without neutrinos or early quintessence,
the amplitude of subhorizon oscillations of photon overdensity,
${\rm Amp}[d\sg]$,
exactly equals the primordial photon overdensity,
$d\sg(\t\to0)=d\i$.
By Sec.~\ref{sec_form_mapping},
any other measure of photon perturbations~$d'\sg$ 
that satisfies conditions~I and~II
coincides with~$d\sg$ on super- and subhorizon scales.
Any such measure, therefore, would also detect no radiation driving:
${\rm Amp}[d\sg']=d\sg'(\t\to0)$.\footnote{\lb{compare_rad_sol}
  For comparison, the perturbation of proper density
  in the Newtonian gauge is
  $$
  \fr{\d n\sg}{n\sg} = 3 \fr{\d T\sg^{(\rm Newt)}}{T\sg} 
  = d\i\lf(\fr{2\sin\f}{\f}-\cos\f + \fr{2\cos\f}{\f^2} - \fr{2\sin\f}{\f^3}\rt).
  $$
  For it, 
  ${\rm Amp}[\d n\sg/n\sg] = 3\lim_{\t\to0}\d n\sg/n\sg$.
}

A formalism-independent distinction between physical driving of perturbations
and a gauge artifact can be made by comparing
the studied model to a reference model 
in which the metric becomes homogeneous by reentry.
In the latter model any driving by metric inhomogeneities
at reentry is objectively absent. 
A consistent procedure to achieve homogeneous geometry before reentry
for any given primordial perturbations of arbitrary species
was described in Sec.~\ref{sec_form_mapping}.
In the homogeneous (FRW) metric  
the evolution of photon fluid overdensity
readily follows from eq.\rf{dot_g} with $c_s^2=1/3$ as
\be
d\sg^{({\rm undriven})} = d\i \cos\f.
\lb{d_rad_undriven}
\ee
Thus if in the superhorizon limit 
the photon and metric perturbations
in the photon-dominated and ``undriven'' models are identical,
after horizon entry
the acoustic oscillations in the two models
have equal amplitude.
There is no causal mechanism for a gravitational boost
of the amplitude in the second model.
Therefore, the amplitude should also not be considered boosted 
in the photon-dominated model.

\begin{figure}[t]
\includegraphics[width=8.5cm]{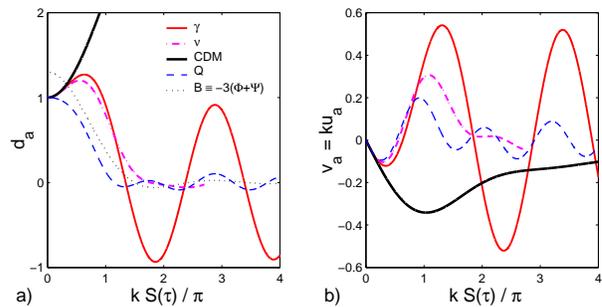}
\caption{{\bf a)} Density perturbations~$d_a$ of typical
cosmological species in the radiation epoch dominated
by tightly coupled photons~($\g$, red solid) and
3~flavors of decoupled relativistic neutrinos~($\nu$, dash-dotted).
The subdominant species are cold dark matter 
(CDM, black solid), 
and a scalar field (tracking quintessence~$Q$, dashed).
The primordial perturbations are assumed adiabatic.
The dotted curve shows the equilibrium value
of $d\sg$ oscillations, eq.\rf{B_def}.
{\bf b)}~Bulk peculiar velocities
         of the same species.
}  
\lb{fig_evol_rad}
\end{figure}

The 3-fold increase of the Newtonian-gauge perturbation
$\d T\sg^{(\rm Newt)}\!/T\sg= \fr13\,d\sg+\Psi$
in the second of the above models occurs 
when the metric becomes homogeneous
on superhorizon scales.
This increase can hardly be attributed to a resonant boost 
of the acoustic oscillations from the decay of their gravitational potential 
at the maximum photon compression, as stated in\ct{HuDodelson02}.
The enhancement of the perturbation measure $\d T\sg^{(\rm Newt)}\!/T\sg$ 
is possible because its evolution 
is not determined only by local dynamics.

\subsubsection{Gravitational impact of perturbations on the CMB and CDM}
\lb{sec_rad_impact}

Fig.~\ref{fig_evol_rad}\,a) displays
the linear evolution of coordinate density perturbations~$d_a(\t,k)$ 
in a photon-baryon fluid, decoupled relativistic neutrinos,
CDM, and a minimally coupled scalar field 
[tracking quintessence, $Q$, with $w_Q(z)= w_{\rm total}(z)$].
Fig.~\ref{fig_evol_rad}\,b) shows 
the corresponding bulk velocities of the species.
The plots describe the radiation epoch, 
dominated by photons and three standard flavors of neutrino.
They are obtained by the numerical integration of 
eqs.\rfs{dotdu_c}{dot_f} and\rfs{Psi_eq}{Phi_eq}
with the regular adiabatic initial conditions,
normalized to $d(k/\H\rightarrow 0)=1$.
The chosen evolution parameter is
a dimensionless phase of photon acoustic oscillations~$kS(\t)$ 
where $S(\t)\equiv \int_0^\t c_s d\t=\t/\sqrt{3}$ (for $R_b\ll1$)
is the size of the acoustic horizon.

In the description of the acoustic dynamics of the CMB by eq.\rf{dot_gk},
the gravitational impact of the species' perturbations 
is mediated through the ``restoring force''
$-c_s^2k^2(d\sg-B)$, with
\be
B(\t,k)=\nbrk{-3(\Phi+\Psi+R_b\Phi)}.
\lb{B_rep}
\ee
The potentials $\Phi$ and $\Psi$ are sourced by perturbations
of densities, peculiar velocities, and anisotropic stress of the species
as described by eqs.\rfs{Psi_eq}{Phi_eq}.

For typical cosmological species,
the growth of subhorizon perturbations 
during the radiation era is insufficient 
for supporting a non-vanishing gravitational potential.
Hence, the equilibrium photon overdensity~$B(\t,k)$
on subhorizon scales tends to zero.
Thus in the radiation era
the gravitational impact of perturbations is confined to horizon entry. 
As for the later subhorizon CMB oscillations,
this impact can only rescale the oscillations' amplitude 
and additively shift the phase.
We now consider the gravitational impact on the CMB
that is caused by decoupled relativistic neutrinos
and a scalar field (early quintessence).

Addition of extra ultrarelativistic neutrinos
changes~$B$ as shown by the solid curve in Fig.~\ref{fig_impact_rad}\,c).
The increase of~$N\snu$ reduces~$B$ during the initial growth of $d\sg$
while enhances~$B$ for a significant period of $d\sg$ fall.
The resulting modulation of the restoring force $-c_s^2k^2(d\sg-B)$ 
somewhat {\it damps\/}~$d\sg$ oscillation.
The reduction of the amplitude of CMB oscillations
by neutrino perturbations\ct{HuSugSmall95,BS} is evident
from the thin solid curve in Fig.~\ref{fig_impact_rad}\,a).

On the contrary, the density perturbation
of canonical tracking quintessence\ct{RatraPeebles88,FerreiraJoyce97,
ZlatevStein_tracking98,OttExpQuint01} 
is significantly reduced right after horizon entry
[see the dashed line in Fig.~\ref{fig_impact_rad}\,a).]
This reduction causes a prominent drop in~$B(\t)$,
shown by the dashed curve in Fig.~\ref{fig_impact_rad}\,c),
at the time when $\dot d_a$ is on average negative.
As a result, the acoustic oscillations gain energy
and their amplitude {\it increases\/}\ct{FerreiraJoyce98}.

The qualitative difference in the evolution of density perturbations 
for quintessence versus the other considered species 
can be understood as follows.
For the regular adiabatic initial conditions, 
the number density perturbations~$d_a$ of all the species 
are equal and the velocities vanish beyond the horizon ($kS\ll 1$).
Moreover, the initial accelerations of all the species, eq.\rf{v_superhor}, 
are also equal, as confirmed by Fig.~\ref{fig_evol_rad}\,b). 
However, the rate of change of~$d_a$ at reentry
is very sensitive to whether or not pressure of the species  
is fully specified by their local density.  
If it is, such as for photons, CDM, and neutrinos,
the ``nonadiabatic'' term on the right-hand side of eq.\rf{dotd_any_gauge} 
vanishes. The corresponding~$d_a$'s agree in their
evolution up to $O(k^4\t^4)$ corrections.
On the other hand, for a classical field 
the nonadiabatic term becomes non-zero during the entry
[\cf the first of eqs.\rf{dot_f}.]
As a consequence,
quintessence density perturbation~$d_Q$ 
[dashed line in Fig.~\ref{fig_evol_rad}\,a)]
noticeably deviates from the perturbations of the other species
already in $O(k^2\t^2)$ order.

Quantitatively, a change of $N\snu$ from $3$ to~$4$
changes the amplitude of $d\sg$ oscillations by $-1.7\%$.  
Addition of equal energy in the form of tracking quintessence,
on the contrary, enhances the amplitude by $1.6\%$.

Perturbations of CDM evolve according to eqs.\rf{dotdu_c} as
\be
d_c = - \int (\Nbi v^i_c)\,d\t, \qquad
  av^i_{c} = - \int (a\Nbi\Phi)\,d\t.
\ee
The impact of metric perturbations on CDM 
is thus determined by a normalized quantity $-(a/a_{\rm ent})\Phi$,
where $a_{\rm ent}\equiv a|_{k\t=1}$.
Fig.~\ref{fig_impact_rad}\,d) shows how this quantity
is affected by the addition of extra neutrinos and quintessence.
By comparing 
Fig.~\ref{fig_impact_rad}\,b) and Fig.~\ref{fig_impact_rad}\,a),
we see that the signs of the gravitational impact
of perturbations in either neutrinos\ct{HuSugSmall95,BS} 
or quintessence\ct{Caldwell03} on the power of CDM
are {\it opposite\/} to the signs 
of the impact of these species on the acoustic amplitude of the CMB.
Thus on the scales entering the horizon before matter-radiation equality 
($\ell\gtrsim200$, $k\gtrsim 10^{-2}$\,Mpc$^{-1}$)
the ratio of CMB to CDM powers,
which is insensitive to the unknown primordial power 
of cosmological perturbations,
is slightly decreased by decoupled particles.
On the other hand, this ratio is
increased by tracking quintessence.

The gravitational impact of perturbations
on the phase of CMB oscillations is easier to analyze in real space\ct{Magueijo92,Baccigalupi98,BB_PRL,BB}.
The phase shift, along with all the other 
characteristics of subhorizon oscillations, 
is fully specified by the singularity of the 
CMB Green's function at the acoustic horizon\ct{BS}.  
An analysis of this singularity\ct{BS} shows that
for adiabatic initial conditions
the phase is affected only by the perturbations 
which propagate {\it faster\/} than the acoustic 
speed of the photon fluid.

Perturbations of both ultrarelativistic neutrinos and 
quintessence propagate at the speed of 
light~$c$, {\it i.e.\/}, faster than the acoustic speed 
$c_s\le c/\sqrt3$.
The phase shift induced by neutrinos
displaces all the CMB peaks in the temperature and E-polarization
spectra by an about equal multipole number\ct{BS}
\be
\D l\snu \approx -57\, \fr{\rho\snu}{\rho\sg+\rho\snu}.
\ee
The three standard types of neutrino give $\D l_{3\nu}\approx -23$,
large enough to be observable with the present 
data\ct{Crotty_Nnu03,Pierpaoli_Nnu03,Hannestad_Nnu03,
        Barger_Nnu03,Trotta:2004ty,Bell:2005dr,WMAPSpII}.
One additional fully populated type of relativistic neutrino 
increases the shift by\ct{BS} $\D l_{+\nu}\simeq -3.4$.  
According to numerical integration of eqs.\rfs{dotdu_c}{dot_f} 
and\rfs{Psi_eq}{Phi_eq},
the same additional radiation density ($\D N_{\rm eff}=+1$) 
in tracking quintessence leads to 
an almost triple shift $\D l_Q\simeq -11$.

\begin{figure}[t]
\includegraphics[width=8.5cm]{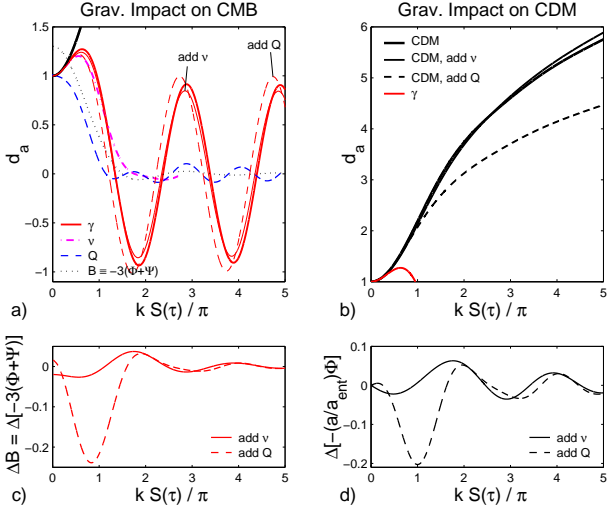}
\caption{RADIATION ERA. 
The impact of perturbations
in decoupled neutrinos and quintessence
on the CMB and CDM.
The wide solid curves give $d\sg$ ({\bf a}) and $d_{\rm CDM}$ ({\bf b}) 
in the epoch dominated by tightly coupled photons and
three standard flavors of decoupled neutrinos.
The thin solid curves describe 
a model with doubled neutrino energy density.
The dashed curves describe a model with the standard neutrino content
plus the same energy density in tracking quintessence,
with $w_Q=1/3$.
The bottom panels demonstrate how these modifications
of the fiducial model change the gravitational terms which drive 
the CMB ({\bf c}) and CDM ({\bf d}).
}  
\lb{fig_impact_rad}
\end{figure}

While in a formal treatment of Boltzmann hierarchy
for the perturbations of free-streaming neutrinos, 
\eg\ct{Hannestad_strongly04}, 
the anisotropic stress might appear
to be the primary cause of their characteristic gravitational imprints,
this conclusion is misleading.
For example, anisotropic stress of minimally coupled tracking quintessence,
whose perturbations also change the amplitude and phase 
of acoustic oscillations, is identically zero.
In particular, a real space approach\ct{BS} shows that
for adiabatic perturbations the phase is shifted
{\it if and only if\/} some perturbations physically propagate 
faster than the acoustic speed~$c_s$.

\subsection{Matter era}
\lb{sec_mat}

\begin{figure}[t]
\includegraphics[width=8.5cm]{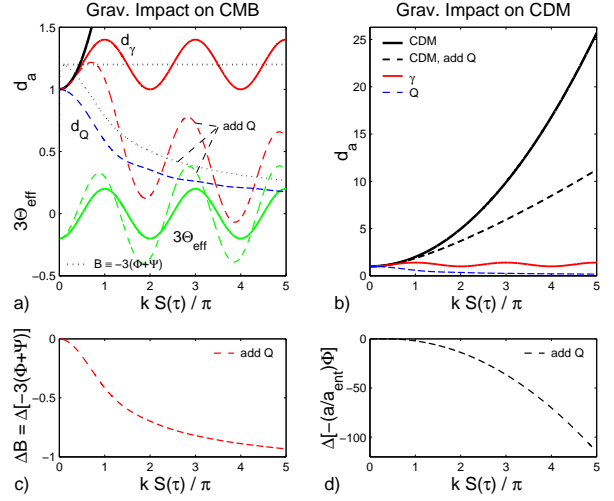}
\caption{MATTER ERA.
The perturbations of the CMB density~$d\sg$ 
and effective temperature~$\Teff$
({\bf a}) and of CDM density~$d_c$ ({\bf b}) 
in the matter era dominated by pressureless matter (solid curves) 
and by equal densities of pressureless matter 
and tracking quintessence, with $w_Q=0$ (dashed curves).
Similarly to Fig.~\ref{fig_impact_rad},
the bottom panels show the change 
induced by quintessence in the gravitational terms 
which drive the CMB ({\bf c}) and CDM ({\bf d}).
}  
\lb{fig_impact_mat}
\end{figure}

\subsubsection{25-fold Sachs-Wolfe suppression}
\lb{sec_SW}

In the epoch dominated by pressureless CDM and baryons 
which have decoupled from the CMB,
metric perturbations (gravitational potentials)
do not decay on subhorizon scales.
Through the finite potentials,
matter continues to influence 
CMB long after horizon entry.
For the perturbations entering the horizon in this epoch,
matter overdensity grows linearly  
in the scale factor $\nbrk{a\propto\t^2}$,
as $d=d\i(1+k^2\t^2/30)$.
If the initial conditions are adiabatic,
photon and neutrino perturbations
contribute negligibly to inhomogeneities of the metric
in the matter era.
Then Newtonian gravitational potentials, 
induced entirely by perturbations of nonrelativistic matter,
remain time-independent
throughout all the linear evolution:
\be
\Phi=\Psi=-\fr15\,d\i,
\lb{Phi_MD}
\ee
as can be verified with eqs.\rf{Psi_eq} and\rf{dotdu_c}.

The acoustic dynamics
and subsequent free-streaming of CMB photons
in the gravitational wells of non-relativistic matter
are well understood on subhorizon scales,
\eg\ct{BondEfstat_CMB87,HuSugSmall95,HuNature,
HuDodelson02,Dod_book,Mukhanov_book}.
However, the traditional description
of horizon entry in the matter era,
\eg\ct{HuSug_toward94,HuSugAnalyt95,HuNature,HuDodelson02},
requires corrections.

For the considered perturbations,
entering the horizon after CMB last scattering but long before 
dark energy domination,
the Doppler and polarization contributions
to the observed CMB anisotropy $\D T/T$
[second and third terms in eq.\rf{los_zeta}]
are negligible.
The ISW contribution
$\dot\Phi+\dot\Psi$ is also absent.
Indeed, the potentials are static
during the matter era and their later decay
due to the dark energy domination is for these modes
adiabatically slow.
The remaining contribution to the
observed $\D T/T$\rf{los_zeta}
is then described by the effective temperature 
perturbation~$\Teff$\rf{Teff_def},
\be
{\D T \ov T}= \Teff = \lf.{\d T \ov T}\rt|_{\rm in}+2\Phi.
\lb{SW}
\ee
Here, we defined
\be
\lf.{\d T \ov T}\rt|_{\rm in} \equiv \fr13\,d_{\g,\,\rm in},
\lb{dT_in}
\ee
with $d_{\g,\,\rm in}\equiv d_{\g}(k\ll\H(\t))$ 
and set $\Phi=\Psi$ in the matter era.
The quantity\rf{dT_in} is a natural measure of
the intrinsic CMB temperature perturbation on superhorizon scales. 
This is the perturbation 
to be always observed for the decoupled CMB photons 
if the metric becomes homogeneous on superhorizon scales
and remains such during the subsequent evolution.

If the initial conditions are adiabatic (Sachs-Wolfe case)
then, by eq.\rf{Phi_MD},
$\Phi=-\fr15 d\i=-\fr15 d_{\g,\,\rm in}$.
Thus the classical Sachs-Wolfe result
$\D T/T=\Phi/3$\ct{SachsWolfe67} is due to
a partial cancellation 
of the primordial CMB density perturbation
$\lf.{\d T / T}\rt|_{\rm in}=-\fr53\,\Phi$
against the gravitational redshift~$2\Phi$
in proportion~$-5:6$. 
The 5-fold suppression of the primordial temperature perturbation
results in an astonishing (and experimentally evident)
suppression of the angular power of CMB temperature~$C_l$ 
by a factor of~$25$.

The traditional interpretations of the Sachs-Wolfe result
also note certain, 2-fold rather than 5-fold, 
suppression of the intrinsic CMB temperature fluctuations 
by the matter potential.
The relation $\D T/T=\Phi/3$
is usually interpreted as 
a partial cancellation between the ``intrinsic'' temperature anisotropy 
$\d T^{\rm (Newt)}/T|_{\rm in}=-\fr23\Phi$ and the 
gravitational redshift~$\Phi$\ct{HuSug_toward94,HuSugAnalyt95,HuNature,HuDodelson02}:
\be 
\qquad\qquad
{\D T \ov T}
     = \lf.{\d T^{\rm (Newt)} \ov T}\rt|_{\rm in}\!\! 
                     + \Phi \quad (\rm misleading!)
\lb{SW_mislead}
\ee
Let us present two examples where such quasi-Newtonian 
considerations lead to wrong expectations.

One example is a classical scenario of ``isocurvature'' initial
conditions, \eg, from a phase transition early in the 
radiation era, when $\rho\sg \gg \rho_c$.  
If immediately after the phase transition $\d\rho\sg+\d\rho_c=0$ 
then $|\d\rho\sg/\rho\sg| \ll |\d\rho_c/\rho_c|$.
For a fixed magnitude of cosmological inhomogeneities
$|\d\rho_c/\rho_c|\sim 10^{-5}$,
the primordial photon perturbations $\d\rho\sg/\rho\sg$
can then be ignored.
However, it would be wrong to conclude
that $\nbrk{\d T^{\rm (Newt)}/T|_{\rm in}=0}$ 
and that from eq.\rf{SW_mislead},  
due to the gravitational redshift in the matter era, 
$\lf.{\D T/T}\rt|_{\rm isoc} = \Phi$ (incorrect!).
The correct answer is $\lf.{\D T/T}\rt|_{\rm isoc}=2\Phi$,
in agreement with eq.\rf{SW}
where now $\lf.{\d T / T}\rt|_{\rm in}=0$.

Another example where the interpretation\rf{SW_mislead}
misleads is the opposite case of
non-zero $\d\rho\sg/\rho\sg$ but homogeneous and isotropic
geometry in the matter era.
If the primordial perturbations of radiation
in this scenario are identical to the perturbations
in the adiabatic model
then the interpretation\rf{SW_mislead} suggests
that now, when the ``2-fold'' suppression by $\Phi$ is absent,
$\lf.{\D T/T}\rt|_{\Phi=0}=-2\lf.{\D T/T}\rt|_{\rm adiab}$ (incorrect!).
The correct answer, immediate from eqs.\rf{SW} and\rf{dT_in}, is 
\be
\lf.{\D T \ov T}\rt|_{\Phi=0}=
  \lf.{\d T \ov T}\rt|_{\rm in}= -5\lf.{\D T \ov T}\rt|_{\rm adiab},
\ee
in agreement with our claim of 5-fold suppression
of $\D T/T$ by $\Phi$ in the adiabatic case.

The homogeneous geometry of the last scenario could be generated
with the matter initial conditions $\nbrk{d_{m,\,\rm in}\approx 0}$,
yet there is no physical reason to expect 
such initial conditions\ct{LindeMukh_TS05}.
However, $d_c$ and the matter-induced potentials $\Phi$ and $\Psi$
may be partly smoothed early during the horizon entry
by non-standard dynamics of the dark sector.
In the limiting case of a perfectly homogeneous metric
since very early time,
we recover the unsuppressed CMB oscillations, 
with the amplitude $d_{\g,\,\rm in}$ in overdensity and 
$\D T/T=\lf.{\d T / T}\rt|_{\rm in}=-5\lf.{\D T/T}\rt|_{\rm adiab}$
in temperature.

\subsubsection{Implications}
\lb{sec_implic}

The observed CMB multipoles with $\ell\lesssim 200$ 
are primarily generated by
the perturbation modes which enter the horizon during the matter era.  
These modes evolve in a static potential of collapsing CDM and baryons. 
The potential suppresses the corresponding $C_l$'s by an order of magnitude,
see the solid curve in Fig.~\ref{fig_cl_ratio}
for the ``concordance'' $\Lambda$CDM 
model\ct{WMAPSpergel,Tegmark03_constr,Seljak05_constr,WMAPSpII}.

The suppression is somewhat reduced for $\ell\lesssim 10$
because of the modes which enter the horizon
when dark energy becomes dominant 
and the potential decays.
For these modes 
the enhancement of CMB power is fully described 
by the ISW term $\dot\Phi+\dot\Psi$, eq.\rf{los_zeta}.
In general, however, 
the reduction of 
the 25-fold Sachs-Wolfe suppression
is not equivalent to the ISW enhancement.
An extreme example is the scenario in which $\Phi+\Psi$
vanishes prior to CMB decoupling.
Given equal primordial perturbations,
the observed CMB power~$C_l$ in this scenario exceeds
the Sachs-Wolfe power by 25~times,
although in both scenarios the ISW effect is absent.

\begin{figure}[t]
\hspace{-0.6cm}\includegraphics[width=6cm]{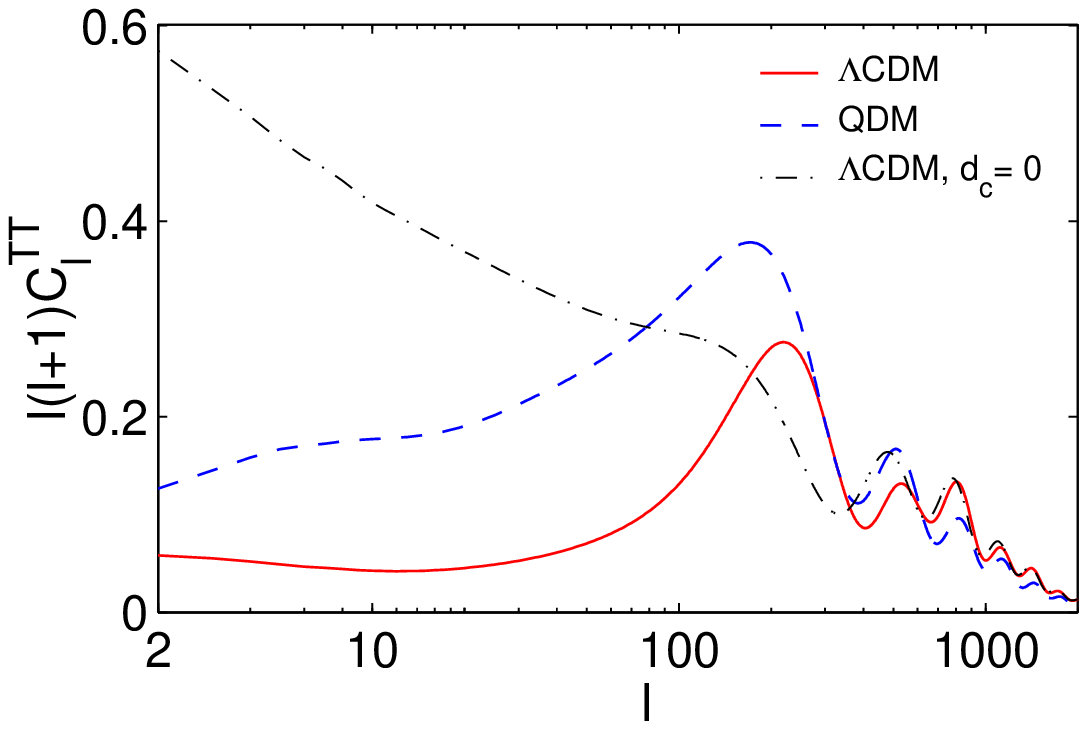}
\caption{
  Suppression of the angular power spectrum
  of CMB temperature~$C_l$  
  for $\ell\lesssim200$ by dark matter perturbations.  
  The figure compares three models which 
  differ only by perturbations in the dark sector:
  The ``concordance'' $\Lambda$CDM 
  model\ct{WMAPSpergel,Tegmark03_constr,Seljak05_constr,WMAPSpII} (solid);
  a model where CDM and cosmological constant are replaced by
  a dynamical scalar field with the same background~$\rho(z)$
  (dashed, the overall power of perturbations 
   is decreased to match the concordance $C_{300}$);
  finally, a $\Lambda$CDM model where CDM density 
  is artificially arranged to be unperturbed beyond the horizon 
  (dash-dotted, the primordial power of photon, baryon, 
   and neutrino perturbations is the same as in the concordance model).
}
\lb{fig_cl_ratio}
\end{figure}

More realistically,
the density perturbation $d_a$ of some of the dominant species
in the matter era may grow 
slower than $[a^2(\rho+p)]^{-1}\sim 1/\H^2\sim \t^2$.
Then, by eq.\rf{Psi_eq}, 
these species contribute only decaying terms
to the gravitational potentials. 
This diminishes the suppression 
of the CMB power at $\ell\lesssim 200$.
As an illustration,
the upper solid and dashed curves in Fig.~\ref{fig_impact_mat}\,a)
show $d\sg$ in a CDM-dominated universe (Sachs-Wolfe case)
and in a model where the energy density of the dark sector 
is split equally between CDM  and quintessence whose 
background density tracks CDM.
In the second model the CMB temperature perturbations
on subhorizon scales are larger by~$93\%$.\footnote{
  If in the matter era some of the dominant species 
  do not cluster, the gravity-driven growth of CDM 
  inhomogeneities also significantly weakens.  This is seen distinctly 
  in Fig.~\ref{fig_impact_mat}\,b).  A classical example of such species
  is baryons prior to their decoupling from the CMB, 
  {\it e.g.\/} Ref.\ct{HuSugSmall95}.
}

Thus any physics which reduces metric inhomogeneities
in the matter era strongly influences CMB temperature anisotropy. 
Some of the influence may be due to a different ISW effect
from the temporal change of the gravitational potentials.
It can also be due to reduced gravitational lensing of the CMB, 
appearing in the second perturbative order.
However, the most prominent impact is caused by 
the very reduction of the potentials,
suppressing the auto-correlation of CMB temperature 
by a factor of~25.

Two examples are shown in Fig.~\ref{fig_cl_ratio}.
The figure displays the CMB temperature power spectra~$C_l$
obtained with a modified \cmbf\ code\ct{CMBFAST96} for three models 
which differ only by the dynamics of {\it perturbations in the dark sectors\/}.
The solid curve describes the concordance 
$\Lambda$CDM model\ct{WMAPSpergel,Tegmark03_constr,Seljak05_constr,WMAPSpII}.
The dashed curve refers to a model where the dark sector 
is composed entirely of a classical canonical scalar field~$Q$.
In this model at every redshift
the field background density $\rho_Q(z)$
is chosen to equal the combined CDM and vacuum 
densities $\rho_c(z)+\rho_\Lambda$ of the $\Lambda$CDM model.
However, since the field perturbations do not grow, 
the metric inhomogeneities of the second model
decay after the horizon entry in the matter era.
As a result, its CMB power for $\ell\lesssim 200$ is enhanced.

The last, dash-dotted, curve in Fig.~\ref{fig_cl_ratio}
describes a model which has the matter content 
of the concordance $\Lambda$CDM model.
However,
while in the previous two scenarios the initial conditions were adiabatic,
here CDM density perturbation 
$d_c$ is set to zero on superhorizon scales
(the initial values for $d\sg$, $d\snu$, and $d_b$ are unchanged).
Then, again, the metric inhomogeneities in the matter era are reduced.
While this model has the least ISW effect,
the smoother metric causes the largest enhancement of
the CMB power among the compared models for $\ell\lesssim 200$.

\section{Summary}
\lb{sec_concl}

There is a variety of 
descriptions of inhomogeneous cosmological evolution
which are apparently dissimilar 
on the scales exceeding the Hubble scale.
To single out the descriptions least afflicted by gauge artifacts,
we suggest considering the measures of perturbations 
that meet the following criteria:
I.~The measure's dynamics is determined entirely by
   the physics within the local Hubble volume.
II.~The measure is practically applicable in Minkowski limit
    and retains its physical meaning in any FRW geometry on all scales.

We showed (Sec.~\ref{sec_phase_sp}) that in the fully nonlinear theory
these criteria are fulfilled for a perturbation of one-particle phase-space 
distribution $f(x^\mu,P_i)$, 
considered as a function of 
particle canonical momenta~$P_i$.
The often used perturbations of $f(x^\mu,\p)$ or $f(x^\mu,\q)$,
where $\p$ is the proper momentum and $\nbrk{\q\equiv a\p}$, 
do not satisfy requirement~I.
Perturbed nonlinear dynamics of radiation 
can be appropriately described by the intensity 
perturbation~$\di(x^\mu,n_i)$,
constructed by integration of $f(x^\mu,P_i)$, eqs.\rf{I_def_a} and\rf{di_def}.
Finally, a description of
perturbed linear dynamics of arbitrary species 
satisfies criteria~I and~II 
if the
perturbations are quantified by coordinate number density perturbations 
$d=\d n_{{\rm coo}}/n_{{\rm coo}}$, eq.\rf{d_a_def}.
In linear theory these measures of perturbations in
phase-space distribution, intensity, and density
are interrelated with eqs.\rf{df2d} and\rf{di2d}.

Similarly to the traditional proper measures of perturbations, 
\eg\ $\D T(\n)/T$ and $\d\rho/\rho$,
the proposed ''coordinate'' measures 
$\di(n_i)$ and $d=\d n_{{\rm coo}}/n_{{\rm coo}}$ 
are gauge-dependent.
Yet the gauge dependence of the proposed,
but not traditional, measures
disappears in the superhorizon limit.
More generally, we prove that on superhorizon scales
the values of the density perturbations 
are the same in terms of any measure satisfying the formal versions 
of criteria~I and~II,
namely, conditions~I$'$ and~II$'$ on page~\pageref{modif_conds}.

In the Newtonian gauge\rf{Newt_gauge_def} 
the suggested variables 
provide additional technical advantages.
One is the explicit Cauchy structure of the dynamical equations
(eq.\rf{dotd_any_gauge}, Sec.~\ref{sec_veloc}, and
Appendix~\ref{sec_dyn}).
Another is a nonsingular superhorizon limit
of the generalized gravitational Poisson equation (Sec.~\ref{sec_metric}).
Analytical solutions in the new formulation inevitably appear 
more tractable\ct{BS}.
As an example, compare the solutions for density
perturbations in the photon-dominated model,
eq.\rf{d_rad} versus the traditional solution in Footnote~\ref{compare_rad_sol}.

In the proposed class of formalisms, 
classical linear density perturbations of individual species
manifestly decouple gravitationally on superhorizon scales.
This view of linear superhorizon evolution of multiple species 
(Sec.~\ref{sec_evol_superhor})
agrees with the picture provided by the conserved curvatures 
of Wands et al.\ct{zeta_a}.
The density perturbations of the species 
recouple during the horizon entry (Sec.~\ref{sec_evol_entry}).

On the other hand, our description reveals
a different from the traditionally assumed
origin of several prominent features in
the angular power of CMB temperature~$C_\ell$.
First of all, the apparent ``resonant boost'' 
of CMB perturbations for $\ell\gtrsim 200$,
presumably by their self-gravity during the horizon entry 
in the radiation era\ct{HuSugSmall95,HuNature,HuFukZalTeg00,HuDodelson02,Dod_book,Mukhanov_book},
is found to be a {\it gauge artifact\/} of the Newtonian gauge. 
We demonstrated (Sec.\ref{sec_rad_driv}) that 
the driving is absent when density perturbations are 
quantified by {\it any\/} measure satisfying conditions~I$'$ and~II$'$.

We showed (Sec.~\ref{sec_rad_driv}) 
that the ``radiation driving'' picture
leads to incorrect expectations for the observed 
CMB signatures of dark species.
Since the resonant self-boost of CMB perturbations
does not physically exist and cannot be
untuned by the gravitational influence of
non-photon species present in the radiation era 
with non-negligible densities,
the CMB signatures of such species are not anomalously amplified.  
This corrected expectation is confirmed by calculations 
for decoupled relativistic neutrinos and for tracking quintessence
(Ref.\ct{BS} and Sec.~\ref{sec_rad_impact}).

A moderate impact of such species on the CMB power
at small scales, less affected by cosmic variance,
nevertheless already allows meaningful constraints
on the species' abundance and nature,
Refs.\ct{Crotty_Nnu03,Pierpaoli_Nnu03,Hannestad_Nnu03,
         Barger_Nnu03,WMAPSpII,Hannestad_strongly04,
                                    Trotta:2004ty,Bell:2005dr}. 
The constraints will significantly improve in the near 
future\ct{Lopez98,Bowen01,BS},
when the smaller angular scales become resolved.

Notably, the CMB is sensitive to 
the propagation velocity of small-scale inhomogeneities
in the dark sectors\ct{Erickson01}.
For neutrinos, quintessence, and any other species
for which this velocity exceeds the sound speed
in the photon fluid~$c_s\approx c/\sqrt3$,
the perturbations cause a nondegenerate uniform
shift of the acoustic peaks\ct{BS,BRev}.
The CMB power is also sensitive to nonadiabatic pressure,
generally developing for a classical field.
As discussed in Sec.~\ref{sec_rad_impact}, 
under adiabatic initial conditions,
the nonadiabatic pressure affects the evolution of species' overdensities
already at the order $O(k^2\t^2)$ 
while the regular pressure enters only at $O(k^4\t^4)$.
In the example of tracking quintessence, 
nonadiabatic pressure causes early suppression 
of quintessence density perturbations
(Sec.~\ref{sec_rad_impact} and Fig.~\ref{fig_evol_rad}).
This enhances the CMB power while neutrino perturbations reduce it
(Fig.~\ref{fig_impact_rad}).

Although CMB perturbations 
are not resonantly boosted on small scales,
the CMB multipoles with $\ell\lesssim 200$ 
are significantly {\it suppressed\/} by the gravitational potential 
which {\it does not decay\/} after the horizon entry {\it in the matter era\/}.
The contribution of the perturbations which enter the horizon
when $|\d p/\d\rho|\ll 1$ and the potential is 
time-independent,
to the angular power of CMB temperature
is suppressed by 25~times (Sec.~\ref{sec_SW}).
The suppression is thus much more prominent
than a factor of~4 reduction of the CMB power 
in the conventional 
interpretations\ct{HuSug_toward94,HuSugAnalyt95,HuNature,HuDodelson02}
of the Sachs-Wolfe effect\ct{SachsWolfe67}.\footnote{
  The CDM-induced suppression of the CMB power at low~$\ell$'s 
  is complimentary to another characteristic impact of 
  dark matter potential 
  on the CMB known as ``baryon drag'' or
  ``baryon loading''\ct{HuSugAnalyt95,HuSugSmall95}.
  The baryon loading is the enhancement of the odd 
  acoustic peaks in $C_\ell$ 
  relative to the even ones
  due to the shift of the 
  zero of $\Teff$ oscillations by $-R_b\Phi$,  
  eqs.\rf{Teff_def} and\rf{B_def}.
  The alternation of the peak heights
  is generally absent in CDM-free models, \eg\ct{McGaughMOND03}.
  While the suppression of $C_l$ at low~$\ell$'s is controled entirely
  by the metric inhomogeneity~$\Phi$,
  the baryon loading is, in addition, proportional to
  baryon density~$\rho_b\propto \Om_bh^2$.
}

As a consequence, 
any mechanism which smoothes metric inhomogeneity
after radiation-matter equality at redshift $\nbrk{z\sim 3500}$ 
significantly enhances CMB anisotropy (Sec.~\ref{sec_implic}).
Examples of such mechanisms are
a noticeable fraction of quintessence at all 
post-equality redshifts, interaction or unification of 
dark matter and dark energy,
\eg\ct{Wetterich88,PerrottaBaccig02,
FarrarPeebles04,Catena:2004ba,Scherrer_kess},
or alternative explanations for 
the apparent signatures of nonbaryonic matter,
\eg\ MOND-inspired theory of gravity\ct{BekensteinMOND04}.\footnote{
  After the first appearance of this work (astro-ph/0405157)
  the predicted enhancement of CMB power spectrum at low $\ell$'s for 
  the Bekenstein T$e$V$e$S\ct{BekensteinMOND04}  
  implementation of MOND\ct{MilgromMOND83}
  has been observed in a numerical analysis of 
  Skordis {\it et al.}\ct{SkordisEtalMOND05}.
  The authors of this analysis, nevertheless, 
  ascribe the enhancement entirely to the ISW effect.
  This interpretation is generally incorrect.
  If by CMB last scattering the metric 
  is perturbed less than in 
  the standard $\Lambda$CDM model,
  the ISW effect in the models 
  with enhanced CMB power
  may be {\it smaller\/} than in the standard model.
}
The enhancement is accounted for by the ISW effect 
when the metric is smoothed after 
CMB last scattering at $z\sim 1100$.
However, even if the ISW effect is absent,
but the metric at or before CMB last scattering 
is already smoother than in the standard scenario,
the enhancement exists.
Thus the experimentally evident suppression 
of CMB multipoles with $\ell\lesssim 200$ 
severely constrains the mechanisms 
which reduce metric perturbations 
at any time in the matter era.

\acknowledgements

This work greatly benefited from the suggestions of
E.~Bertschinger, A.~Friedland, A.~Guth, S.~Habib, M.~Martin, 
N.~Padmanabhan, M.~Perelstein, U.~Seljak,
and an anonymous referee of Phys.~Rev.~D.
Part of the work was performed under 
financial support of Princeton University.
This paper was written under the auspices 
of the National Nuclear Security Administration 
of the U.S. Department of Energy at 
Los Alamos National Laboratory under Contract No. DE-AC52-06NA25396.

\appendix

\section{Perturbative Dynamics of Typical Species}
\lb{sec_dyn}

In this Appendix we give the closed equations
for the linear evolution 
of scalar perturbations of typical cosmological species:
cold dark matter~($c$),
decoupled neutrinos~($\nu$),
photon-baryon plasma~($\g b$),  
and a classical scalar field, or quintessence,~($\f$).
We use the Newtonian gauge\rf{Newt_gauge_def},
in which the gravitational potentials $\Phi$ and $\Psi$ are determined by
the perturbations of the total density, velocity, and anisotropic stress
with eqs.\rfs{Psi_eq}{Phi_eq}.

\subsection{Cold dark matter (CDM)} 
\lb{sec_CDM}

Before CDM streams cross, in particular, in the linear regime, 
CDM is equivalent to a fluid with vanishing
pressure and vanishing anisotropic stress ($p_c=\d p_c=0$, $\s_c=0$).
Then from eqs.\rf{dotd_any_gauge} and\rf{dot_u}
\be
\dot d_c = \Nb^2 u_c,\qquad
\dot u_c = - \H u_c+\Phi.
\lb{dotdu_c}
\ee
Elimination of $u_c$
gives a single
second-order equation for CDM density perturbation
\be 
\textstyle
\ddot d_c + \H\dot d_c
          = \Nb^2\Phi.
\lb{dot_c}
\ee

\subsection{Neutrinos}
\lb{sec_neutr}

\subsubsection{General case}

When the universe cools below~$1$\,MeV, 
weak interactions of the standard neutrinos
become cosmologically irrelevant.
Then the full evolution of neutrino phase-space distribution~$f$
is governed by collisionless Boltzmann equation\rf{df_dot_free}.
Straightforward linearization of eq.\rf{df_dot_free} 
shows that the scalar component 
of $f$ perturbation
\be
df(x^i,P_i)\equiv f(x^i,P_i)-\bar f(P)
\lb{df_def}
\ee
evolves in the linear order in the Newtonian gauge\rf{Newt_gauge_def} as 
\be
\lf(df\rt)\odot
+V_i\,\Nbi\lf(df\rt)
 = E\,\fr{\,\pd\!\bar f}{\pd E}\,V_i\,\Nbi\lf(\Phi+V^2\,\Psi\rt),
\lb{df_dot}
\ee
where $E\equiv (P^2+m^2a^2)^{1/2}$ and $V_i\equiv P_i/E$.

The initial conditions for the perturbation~$df$
on superhorizon scales can be determined as follows.
Consider spatial hypersurfaces of  
constant radiation density~$\rho_r$ and, to be specific, 
assume that neutrino chemical potential is unperturbed.
Then prior to neutrino decoupling
the neutrino phase-space distribution $f\spr(\bm{p}\spr)$ 
in proper variables $d^3\x\spr d^3\bm{p}\spr$ 
is spatially uniform (independent of~$x^i$)
on a hypersurface $\rho_r=\const$.
In a gauge of uniform radiation density 
and zero shear,
in which $\rho_r=\const$ defines the hypersurfaces of constant time 
and the 3-metric is $g_{ij}=a^2(1+2\zeta_r)\d_{ij}$,
the proper and canonical neutrino momenta
are related as $p^i\spr=P_i/[a(1+\zeta_r)]$.
Thus in the variables $(x^i,P_i)$
\be
f(x^i,P_i)=f\spr\lf(\fr{P_i}{a[1+\zeta_r(x^i)]}\rt).
\lb{ffpr}
\ee
(We used that $d^3\x\spr d^3\bm{p}\spr=d^3x^id^3P_i$.)
The corresponding linear perturbation of $f(x^i,P_i)$ is
\be
df = -P\fr{\pd \bar f(P)}{\pd P}\,\zeta_r(x^i). 
\lb{df_ic}
\ee
Note that substitution of the last result in eq.\rf{df2d}
or integration of eq.\rf{ffpr} over $d^3P_i$
returns $d\snu=3\zeta_r$.

Since $f(x^i,P_i)$ is a 4-scalar and
for the decoupled particles $\bar f(P)$ is time-independent,
$df$ is a gauge invariant.
Therefore,
eq.\rf{df_ic} gives the initial conditions
for $df$~evolution in any gauge, 
including the Newtonian.

\subsubsection{Ultrarelativistic limit}
\lb{sec_neutr_relat}

Perturbation of the intensity of ultrarelativistic neutrinos 
is conveniently described by the variable $\di(x^\mu,n_i)$, 
introduced by eq.\rf{di_def}.
After neutrino decoupling,
the full and the linear evolution of neutrino intensity perturbation
is given by eqs.\rf{doti_gen_a} and\rf{dotD} respectively.
The latter shows that for linear scalar modes in the Newtonian gauge
\be
\dot \di\snu + n_i\Nbi \di\snu  
  = -4 n_i\Nbi(\Phi+\Psi).
\lb{dot_di_nu}
\ee

Scalar perturbations of $\di\snu(n_i)$ 
may be parameterized by 
scalar multipole potentials~$d_{\nu l}$ as\ct{BS}
\be\textstyle
\fr34\di(n_i)&\equiv& D(n_i) \,=
\nn\\
&=& \sum_{l=0}^{\infty}(-1)^l\,(2l+1)\,P_l
                \!\lf(n_i\textstyle\fr{\Nbi}{\Nb}\rt)\Nb^l d_l,
\lb{dl_def}
\ee
where $P_l(z)$ are the Legendre polynomials. 
The potentials~$d_{\nu l}$ with $\ell=0,\,1$, and~$2$ describe 
the neutrino number density perturbation, velocity potential, 
and anisotropic stress potential respectively:
$d_{\nu 0}=d\snu$, $d_{\nu 1}=u\snu$, and
$d_{\nu 2}=\s\snu$\ct{BS}.
In the special case of a plane-wave perturbation
$\di(\x)\propto \exp{(i\k\cdot\x)}$,
$\Nb^l d_l$ are proportional to the 
scalar ($m=0$) multipoles
$\di_{l0}=\int d^2\n\, Y^*_{l0}(n_i)\di(n_i)$,
where the polar direction is taken along~$\k$.
By eq.\rf{dot_di_nu}
and the identity $(2l+1)\mu P_l(\mu)=(l+1)P_{l+1}(\mu)+lP_{l-1}(\mu)$,
the neutrino multipole potentials evolve as
\be\textstyle
\dot d_{\nu l}=\fr{l}{2l+1}\,d_{\nu,l-1}
     +\fr{l+1}{2l+1}\,\Nb^2 d_{\nu,l+1}
     +\d_{l1}(\Phi+\Psi)
\lb{dot_dl}
\ee
($\d_{l1}$ in the last term is the Kronecker symbol.)

\subsection{Photon-baryon plasma}

\subsubsection{Tight coupling limit}

Perturbations in the tightly coupled photon-baryon plasma 
($u\sg=u_b$, $\s_{\g b}=0$) 
propagate as sound waves with an acoustic speed, \eg\ct{Dod_book},
\be
c_s^2 \equiv 
{\d p_{\g b}\ov \d\rho_{\g b}}
={\dot p_{\g b}\ov \dot\rho_{\g b}}
         =\fr1{3(1+R_b)}.
\lb{c_s_def}
\ee
Here $R_b\equiv \rho_b/(\rho_{\g}+p_{\g})=3\rho_b/(4\rho_{\g})$
and the photon-baryon perturbations are assumed internally
adiabatic ($d\sg=d_b$).
In this limit, 
from eq.\rf{dotd_any_gauge} 
and eqs.\rf{dot_u},\rf{d_a_Newt},
\be
 \ba{rcl}
\dot d\sg&=&\Nb^2u\sg, \\
\dot u\sg&=&c_s^2\,d\sg-\H(1-3c_s^2)u\sg+\Phi+3c_s^2\Psi.\vsp
 \ea
\lb{dot_gdu}
\ee
Elimination of $u\sg$ yields
\be
\textstyle
\ddot d\sg + \H(1-3c_s^2)\,\dot d\sg - c_s^2\Nb^2d\sg 
          = \Nb^2(\Phi+3c_s^2\Psi).~~~
\lb{dot_g}
\ee

To describe photon and baryon inhomogeneities beyond the tight coupling 
limit we should confront directional variation of photon intensity
and account for the dependence of photon-baryon scattering 
on photon polarization.

\subsubsection{Photon intensity}
\lb{sec_phot_T}

The scalar component of the perturbation $\di\sg(n_i)$ 
of polarization-summed photon intensity 
evolves according to the transport equation\rf{dot_di_Newt}
with additional Thomson scattering 
terms\ct{Kaiser83,BondEfstathiou84,MaBert95}:
\be
\dot \di\sg \!&+&\! n_i\Nbi \di\sg  
  = -4 n_i\Nbi(\Phi+\Psi) + 
\lb{dot_di_g}\\
    &+&\!\fr1{\t_c}\lf\{ -\di\sg+\fr43\,d\sg+4n_i v_b^i + 
    \lf[(n_i\Nbi)^2-\fr13\Nb^2\rt]\!\pip\rt\}.
\nn
\ee
Here, 
\be
\t_c\equiv ({{an_e\sigma}}_{\!\rm Thomson})^{-1}
\lb{t_c_def}
\ee
is a mean conformal time of a photon collisionless flight,
$v_b^i=-\Nbi u_b$ is the velocity of baryons,
and $\pip$ is defined below by eq.\rf{q_def}.
Similarly to the case of neutrinos,
we introduce scalar multipoles~$d_{\g l}$
of photon intensity perturbation~$\di\sg(n_i)$ 
according to expansion\rf{dl_def}.
Then transport equation\rf{dot_di_g} 
expands into a hierarchy of multipole evolution\footnote{
  In momentum space,
  eqs.\rf{mult_gamma} are related to the analogous 
  Newtonian-gauge equations of Refs.\ct{MaBert95} and\ct{CMBFAST96} 
  by elementary substitutions
  $\d\sg = 4\D_{T0}^{(S)}\to 4(\fr13d\sg+\Psi)$,
  $\theta\sg = 3k\D_{T1}^{(S)}\to k^2u\sg$,
  and $F_{\g l}=4\D_{Tl}^{(S)} \to \fr43k^ld_{\g l}$.
}
\be
 \ba{rcl}
\dot d\sg&=&\Nb^2u\sg, \\
\dot u\sg&=&\fr13\,d\sg+\fr23\Nb^2\s\sg+\Phi+\Psi-\fr1{\t_c}\lf(u\sg-u_b\rt),
 \Vsp\\
\dot \s\sg&=& \fr25\,u\sg 
                +\fr35\,\Nb^2d_{\g,3}
             -\fr1{\t_c}\lf(\s\sg -\fr1{10}\,\pip\rt),
\vsp
 \Vsp\\
\dot d_{\g l}&=& \fr{l}{2l+1}\,d_{\g,l-1}
                +\fr{l+1}{2l+1}\,\Nb^2d_{\g,l+1}-\fr1{\t_c}\,d_{\g l}, 
       ~~l\ge 3,
\Vsp
 \ea
\lb{mult_gamma}
\ee
where for the last equation $d_{\g 2}=\s\sg$.

\subsubsection{Photon polarization}
\lb{sec_phot_P}

The phase-space density of partially polarized photons 
with spacetime position~$x^\mu$ and momentum~$P_i$
is described by a $2\times2$ Hermitian tensor $f_{\al\b}(x^\mu,P_i)$,
whose indexes $\al$ and $\b$ 
refer to two spatial directions orthogonal to~$P_i$.
Given a polarimeter selecting the photons whose
electric field varies as $\bm{E}\propto\en\exp(iP_i x^i)$,
where $\en$ is a complex unit 3-vector orthogonal to~$P_i$
($g_{ij}\eps^{*i}\eps^j=1$ and $P_i\eps^i=0$), 
the tensor $f_{\al\b}$ specifies
the phase-space density of the photons passing the polarimeter 
as $\eps^{*\al}f_{\al\b}\eps^\b$.

We define the intensity tensor
of partially polarized photons by 
integrating $f_{\al\b}(x^\mu,P_i)$ over~$P^3dP$,
similarly to eq.\rf{I_def_a}: 
\be
I_{\al\b}(x^\mu,n_i)\equiv \int_0^{\infty} P^3dP\,f_{\al\b}(x^\mu,n_iP).
\lb{I_pol_def}
\ee
The connection of $I_{\al\b}$ to the energy-momentum tensor
of photons with a specific polarization~$\en$ is given by eq.\rf{Tmunu_int_a}
with $I\to \eps^{*\al}f_{\al\b}\eps^\b$.
Finally, we generalize the intensity perturbation~$\di$ 
of eq.\rf{di_def} to polarized radiation 
as
\be
\di_{\al\b}(x^\mu,n_i)\equiv \fr{I_{\al\b}}{\bar I}-1,
\lb{di_pol_def}
\ee
where $\bar I\equiv \mathop{\rm Tr}\bar I_{\al\b}$ is the 
polarization-summed background intensity.

We will restrict the discussion of polarization evolution 
to harmonic plane-wave perturbations,
which vary in space as $\exp{(i\k\cdot\x)}$.
While the general analysis is also possible,
we defer it to a separate publication.
In linear theory,
solving the evolution of the plane-wave perturbations
is sufficient for constructing
the CMB transfer functions, 
summarized in Appendix~\ref{sec_Cl}.

For the plane-wave perturbations of intensity~$\di_{\al\b}(\n)$,
we use the standard basis vectors 
$\en_\theta$ and $\en_\phi$, both orthogonal to $\n$,
where $\en_\theta$ lies in the ($\k$,\,$\n$) plain  
and $\en_\phi$ is orthogonal to it.
In the basis $(\en_\theta,\en_\phi)$, 
a scalar (axially symmetric about $\k$) 
perturbation $\di_{\al\b}$ 
has the general form\ct{BaskoPolnarev80}
\be
\di_{\al\b}=\fr12\lf(\ba{cc} \di+\dip & 0 \\ 0 & \di-\dip \ea\rt).
\ee
Its Stokes' polarization parameters are 
$Q = \dip(x^\mu,\^\k\,{\cdot}\,\n)$ and $V=U=0$.

The photon polarization~$\dip\sg(x^\mu,n_i)$ evolves according to 
a transport equation\ct{Kaiser83,BondEfstathiou84,MaBert95}
\be
\dot\dip\sg + n_i\Nbi\dip\sg  
  = \fr1{\t_c}\lf\{-\dip\sg + [(n_i\Nbi)^2 -\Nb^2]\pip\rt\}\!.~~
\lb{dot_p}
\ee
In the last terms of eqs.\rf{dot_p} and\rf{dot_di_g}, 
\be
\pip \equiv  \s\sg + \~ p_{\g 0} + p_{\g 2}.
\lb{q_def}
\ee
The variables $\~p_{\g 0}$ and $p_{\g 2}$ are defined by 
the following expansion of $q\sg(n_i)$ into scalar
multipole potentials\ct{BB}:
\be\textstyle
\fr34\dip\sg(n_i)&=& -\Nb^2 \~p_{\g 0}+3(n_i \Nbi)\Nb^2 \~p_{\g 1}\,+
\lb{pl_def}
\\
&&+\ \sum_{l=2}^{\infty}(-1)^l\,(2l+1)\,P_l
                \!\lf(n_i\textstyle\fr{\Nbi}{\Nb}\rt)\Nb^l p_{\g l}.
\nn
\ee
The hierarchy of evolution equations
for the polarization multipole potentials
follows from eq.\rf{dot_p} as\footnote{
  In momentum space, eqs.\rf{mult_gamma_pol}
  are related to the equations of Refs.\ct{MaBert95} and\ct{CMBFAST96} by
  $G_{\g l}=4\D^{(S)}_{Pl}\to \fr43k^{l+2}\~p_{\g l}$
  for $\ell=0,1$ and 
  $G_{\g l}=4\D^{(S)}_{Pl}\to \fr43k^l p_{\g l}$ for $\ell\ge 2$.
}
\be
 \ba{rcl}
\dot{\~ p}_{\g 0}&=&
    \Nb^2 \~ p_{\g 1} - \fr1{\t_c}\lf(\~ p_{\g 0} - \fr12\,\pip\rt),\Vsp\\
\dot{\~ p}_{\g 1}&=& \fr13\,\~p\sg{}_0
                     -\fr23\,p_{\g 2}
                     -\fr1{\t_c}\,\~ p_{\g 1},
 \Vsp\\
\dot p_{\g 2}&=& -\fr25\,\Nb^2 \~p_{\g 1}
                +\fr35\,\Nb^2p_{\g,3}
             -\fr1{\t_c}\lf(p_{\g 2}-\fr1{10}\,\pip\rt),
 \Vsp\\
\dot p_{\g l}&=& \fr{l}{2l+1}\,p_{\g,l-1}
                +\fr{l+1}{2l+1}\,\Nb^2p_{\g,l+1}
                -\fr1{\t_c}\,p_{\g l}, ~~l\ge 3.
 \Vsp
 \ea
\lb{mult_gamma_pol}
\ee

\subsubsection{Baryons}

The evolution equations for baryons are similar to 
eq.\rf{dotdu_c} for CDM.
The growth of the baryon density perturbation
is, again, minus the divergence of their velocity.
The equation for baryon acceleration contains
all the terms of eq.\rf{dotdu_c}
and an additional photon pressure term,
related to the last term 
in $\dot u\sg$ equation\rf{mult_gamma}
by local momentum conservation in photon-baryon
scattering\ct{MaBert95}:
\be
 \ba{rcl}
\dot d_b&=&\Nb^2u_b,\\
\dot u_b&=&-\H\,u_b+\Phi-\fr1{R_b\t_c}\lf(u_b-u\sg\rt).\vsp
 \ea
\lb{mult_bar}
\ee

\subsection{Quintessence} 
\lb{sec_quint}

Quintessence\ct{Wetterich88,RatraPeebles88,FerreiraJoyce97,Caldwell98}
is defined as a classical scalar field 
with canonical Lagrangian density ${\cal L}=K-V(\phi)$, 
where $K\equiv -\fr12g^{\mu\nu}\phi_{,\mu}\phi_{,\nu}$.
In contrast to the previously considered species,
the pressure of a classical field is not a local function of
the field energy density.
As a consequence, nonadiabatic pressure and the right-hand side 
of the density conservation equation\rf{dotd_any_gauge} 
are now not zero.
Nevertheless, if the field and the species which couple to it
are perturbed internally adiabatically
then, according to Sec.~\ref{sec_evol_superhor},
the field density perturbation~$d_\phi$
is time-independent on superhorizon scales.

The evolution of quintessence can be described by the field equation 
$\phi^{;\mu}_{;\mu}=V'(\phi)$,
linearization of which in the Newtonian gauge yields
\be
\d\ddot\phi+2\H\d \dot\phi+(k^2+a^2V'')\d \phi\,=
\lb{dd_phi}
\\
=\,\dot\phi\,(\dot\Phi+3\dot\Psi)-2a^2V'\Phi.
\nn
\ee
The Newtonian potentials $\Phi$ and $\Psi$
are affected by both the field perturbation~$\d\phi$ 
and its rate of change~$\d\dot\phi$
through the field energy-momentum tensor
$T^{(\phi)}_{\mu\nu}=\pd_\mu\phi\pd_\nu\phi+g_{\mu\nu}{\cal L}$.

When the scalar field is the dominant component of the universe,
\eg, during inflation, the Newtonian potentials 
are equal and unambiguously determined by $\d\phi$ and~$\d\dot\phi$.
Then $\d\phi$ and $\Phi=\Psi$ can be elegantly combined
into a Mukhanov-Sasaki variable $v=a(\d\phi+\fr{\dot\phi}{\H}\Phi)$,
obeying a simple closed equation of an oscillator with a varying 
mass\ct{Sasaki_infl86,Mukhanov_infl88,Mukh_Rept}.

However, after reheating 
a scalar field no longer dominates the universe energy density.
In this case the field dynamics can be integrated,
at least numerically, in the Newtonian gauge using the general 
equations\rf{dotd_any_gauge} and\rf{dot_u}.

In the zeroth and linear orders of perturbation theory 
$\rho_\phi=K+V$ and $p_\phi=K-V$, with $K=-\fr12g^{00}\dot\phi^2$,
and $T^0_i=\pd^0\!\phi\,\pd_i\phi$.
Hence, the variables~$(d_\phi,u_\phi)$
are connected with $(\d\dot\phi,\d\phi)$ as
\be
d_\phi=\fr{\d\dot\phi}{\dot\phi}+a^2V'\fr{\d\phi}{\dot\phi^2}-\Phi-3\Psi,
\qquad
u_\phi=\fr{\d\phi}{\dot\phi}.~~~
\ee
Since, by the last equation, $\d p_\phi = \d\rho_\phi - 2\dot V u_\phi$
and since anisotropic stress of 
a minimally coupled scalar field is zero,
eqs.\rf{dotd_any_gauge} and\rf{dot_u} yield:
\be
 \ba{rcl}
\dot d_\phi &=& \Nb^2 u_\phi 
        + {\textstyle \fr{\dot V}{K}}\,\dc_\phi,\\
\dot u_\phi &=& \dc_\phi - \H u_\phi+\Phi.
  \ea
\lb{dot_f}
\ee
Here, $\dc_\phi\equiv d_\phi+3\H u_\phi+3\Psi$ [it equals
$\d\rho_\phi/(\rho_\phi+p_\phi)$ in the gauge comoving to the field].
The ratio $\dot V/K=2a^2V'(\phi)/\dot\phi$ in eq.\rf{dot_f} 
is set by a quintessence background ``equation of state''
$w_\phi\equiv p_\phi/\rho_\phi={K-V \ov K+V}$:
\be\textstyle 
~~\fr{\dot V}{K}= -3\H(1-c^2_{\phi\,\rm ad}),\quad
c^2_{\phi\,\rm ad}\equiv \fr{\dot p_\phi}{\dot\rho_\phi}
   = w_\phi-\fr{\dot w_\phi}{3(1+w_\phi)\H}.
\nn
\ee
Note that the ``adiabatic sound speed'' $c^2_{\phi\,\rm ad}$ 
does not describe any physical propagation.
Inhomogeneities of $\phi$ propagate at the speed 
of light\ct{Caldwell98,HuEisenTegWhite98}, 
as evident directly from the field equation\rf{dd_phi}.

\subsection{All}

In each of the above cases
the perturbative dynamics in the fixed 
Newtonian gauge is formulated as 
an explicit initial-value Cauchy problem. 
For the considered matter content, 
an {\it arbitrary\/} set of initial values for 
($d_c,u_c,d_{\nu l},d_{\g l},p_{\g l},d_b,u_b,d_\phi,u_\phi$)
{\it fully\/} specifies 
physically {\it distinct\/} and consistent 
initial conditions for scalar perturbations.

\section{CMB Transfer Functions and $C_l$'s}
\lb{sec_Cl}

Here we summarize 
the formulas for the angular spectra of CMB 
temperature anisotropy and polarization
produced by scalar perturbations.
To be concrete, we consider adiabatic primordial perturbations
with the primordial power spectrum 
\be
\langle\zeta\i(\k)\,\zeta\i^*(\k')\rangle 
=(2\pi)^3\dd^{(3)}(\k-\k')\,P_{\zeta}(k).
\lb{PS_def}
\ee
Correspondingly,
we normalize the perturbation variables 
in the transfer functions below [eq.\rf{T_TE}]
to $\zeta\i\equiv\lf.\zeta\rt|_{k/\H(\t)\ll1}=1$.
For the general nonadiabatic scenarios, $P_{\zeta}(k)$ would be replaced 
by a matrix of the primordial correlations of all the possible 
scalar modes with the momentum~$k$.

The angular correlation spectra
of the observed CMB temperature and polarization are
\be
C^{XY}_l \!\equiv \langle a^{\!*X}_{lm}a^Y_{lm}\rangle 
   = \int\! \fr{2k^2dk}{\pi}~T^X_l(k)\,T^Y_l(k)\,P_{\zeta}(k).~~~
\lb{Cl_def}
\ee
In this equation the superscripts $X$ and $Y$ denote 
either temperature anisotropy~($T$)
or curl-free polarization~($E$)\ct{Seljak:1996gy,Kamionkowski:1996zd,ZalSel_allsky97},
and $a^{X,Y}_{lm}$ are the corresponding observed multipoles.
The transfer functions $T^X_l(k)$ are given by 
the line-of-sight integrals\ct{CMBFAST96,ZalSel_allsky97}
\be
T^{T,E}_l(k) = \int_0^{\t_0}\!d\t\,
     S^{T,E}(\t,k)\,j_l(k(\t_0-\t)),  
\lb{Ttransf_full}
\ee
where $j_l(x)=(-x)^l\lf(\fr{d}{xdx}\rt)^l\fr{\sin x}{x}$ 
is a spherical Bessel function, 
and the sources of temperature perturbation and polarization 
equal
\be
S^T(\t,k) &=&
        \dot g \lf[\Teff+u_b\,\fr{\pd}{\pd\t_0}
        +\fr14\,{\pip}\lf(\fr{\pd^2}{\pd\t_0^2}+\fr{k^2}3\rt)\rt]+\quad
\nn\\
        &&+\ g\lf(\dot\Phi+\dot\Psi\rt),
\lb{T_TE}
\\
S^E(\t,k) &=& \lf[\fr{(l+2)!}{(l-2)!}\rt]^{1/2}\!\!
              \fr{\dot g\, \pip}{4(\t_0-\t)^2}.
\nn\\ \nn
\ee
In eqs.\rf{T_TE},
$\Teff$ is defined by eq.\rf{Teff_def},
$\pip$ by eq.\rf{q_def}, and
\be
g(\t)\equiv \exp\lf(-\int^{\t_0}_{\t}\!\fr{d\t'}{\t_c}\rt)
\lb{visfun_def}
\ee
is the probability for a CMB photon 
emitted at time~$\t$
to reach the Earth unscattered.


\bibliography{drivbib}

\end{document}